\RequirePackage{fix-cm}
\documentclass[twocolumn]{svjour3}          
\smartqed  
\usepackage{graphicx}
\usepackage{lineno,hyperref}
\usepackage{graphics}
\usepackage{amsmath}
\usepackage{cuted, ragged2e}
\usepackage{multicol,graphicx,float}
\usepackage{enumitem}
\usepackage{lipsum, array, booktabs, caption}
\usepackage{multicol,graphicx,float}
\usepackage[english]{babel}
\usepackage{color}
\usepackage{tabularx}
\usepackage{epstopdf}
\usepackage{multirow}
\usepackage{babel,blindtext}
\usepackage{subcaption}
\usepackage[linesnumbered,ruled,vlined]{algorithm2e}
\usepackage{mathtools}
\usepackage[numbers]{natbib}

\begin{document}

\title{Enhancing chaos in multistability regions of Duffing map for an image encryption algorithm}


\author{Hayder Natiq\and Animesh Roy \and Santo Banerjee \and Amar P. Misra \and N. A. A. Fataf 
}


\institute{Hayder Natiq \at
              Information Technology Collage, Imam Ja’afar Al-Sadiq University, Iraq \\  
           \and
           Animesh Roy \at
          Department of Mathematics, Siksha Bhavana, Visva-Bharati University, Santiniketan, 731 235, India\\
          \and
           Santo Banerjee \at
          Department of Mathematical Sciences, Giuseppe Luigi Lagrange, Politecnico di Torino, Corso Duca degli Abruzzi 24, Torino, Italy\\
          \email{santoban@gmail.com}
          \and 
           Amar P. Misra \at
          Department of Mathematics, Siksha Bhavana, Visva-Bharati University, Santiniketan, 731 235, India\\
          \and 
           N. A. A. Fataf \at
          Centre for Defence Foundation Studies, Universiti Pertahanan Nasional Malaysia, Sungai Besi, Malaysia\\          
}


\maketitle

\begin{abstract}
This paper investigates and analyzes the dynamics of the two-dimensional Duffing map. Multistability behavior has been observed from the system numerically. Such behavior, especially the coexistence of chaotic and periodic attractors, is undesirable in the applications of chaos-based cryptography. Therefore, we design and implement a Sine-Cosine chaotification technique to enhance chaos in the multistable regions. Furthermore, this paper proposes a new image encryption algorithm to examine the performance of the generalized Duffing map in cryptography applications. Simulation results and security analysis reveal that the proposed algorithm can effectively encrypt and decrypt several image types with a high level of security.	

\end{abstract}

\section{Introduction}
\label{intro}
Information security has become crucial due to the rapid development of multimedia and Internet technology, especially images, that are shared online. Recent security analyses revealed that the classic encryption algorithms, such as Advanced Encryption Standard (AES) and Data Encryption Standard (DES), are undesirable for images because of their characteristics \cite{Cao2018}. Therefore, several approaches for increasing security have been proposed, such as cellular automata \cite{Wu2016}, DNA coding \cite{Chai2017}, compressive sensing \cite{Zhang2016,Nan2022}, wavelet transmission \cite{Luo2015}, and chaos \cite{Natiq2022Image,Ibrahim2022}. Due to the high security and fast speed of the chaos-based image encryption technique and the similarity between the features of chaotic systems and images, this technique has become widely used and most preferable in cryptography applications \cite{natiq2019degenerating,natiq2020Ehancing,Saidi2020}.

Many image encryption schemes have thus been developed using chaotic systems  \cite{liao2010novel,wu2012image,zhou2013image,xu2016novel,cao2018novel,cao2020designing,Shen2022,Kumar2022}. It has been established that the security level of a chaos-based encryption algorithm is highly dependent on the characteristics of the employed chaotic maps \cite{alvarez2006some}. However, recent investigations have revealed that numerous employed maps can have some drawbacks, such as chaos degradation with finite precision platforms, low complex performance, narrow and discontinuous chaotic ranges \cite{hua2017sine}. In this way, several studies have been  performed to improve the characteristic of chaotic maps by proposing different chaoticfication techniques. For example, Hua et al. \cite{hua20152d,hua2016image} enhanced the chaotic behaviors of the Logistic map by modulating its output using a nonlinear transforme. Natiq et al. \cite{Natiq2018} enhanced the chaos complexity of the 2D Henon map using a Sine map for image encryptions. Hua et al. \cite{hua2019cosine} proposed a Cosine chaoticfication technique to generate robust chaotic maps for encrypting images. 

However, some investigations on the dynamics of chaotic systems have discovered appealing nonlinear phenomena, namely multistability behaviors or coexisting attractors \cite{natiq2018self}. Multi-stable chaotic systems can exhibit more than one chaotic or periodic attractor with appropriate initial conditions. Multi-stable chaotic systems with continuous-time have been presented during the last few years \cite{rahim2019dynamics}. Meanwhile, little attention has been focused on the multistability in discrete-time chaotic systems or maps \cite{natiq2019can}. Note that a multi-stable system with coexisting chaotic and periodic attractors is not preferable in cryptography applications. Therefore, it is crucial to determine either the regions of coexisting only chaotic attractors or a suitable chaotification technique to enhance chaos in the multistability regions.

\par
In this work, we revisit the dynamics of a 2D discrete chaotic system, namely, the 2D-Duffing map. The numerical investigations show that the Duffing map can produce multistability behaviors in which the coexistence of chaotic and non-chaotic attractors and the coexistence of two chaotic attractors can observe with a specific set of system parameters. It is imperative to note that such complicated behavior is rare in low-dimensional chaotic maps. Therefore, we introduce a chaotification technique based on two trigonometric functions to overcome this situation. Dynamical properties show that the proposed chaotification technique can improve the chaotic and non-chaotic attractors to become hyperchaotic attractors. Besides, it enhances the unpredictability and randomness of the map. Based on the generalized Duffing map, we propose a new image encryption algorithm with the principles of confusion and diffusion. Firstly, the hyperchaotic sequences are generated for scrambling of plain-image pixels. Then, the diffusion process is accomplished  by the elliptic curves, S-box, and hyperchaotic sequences. The simulation results confirm that the proposed encryption algorithm can effectively encrypt various kinds of digital images including Grey-scale, RGB, medical, and hand writing images. Security analysis shows that this algorithm can resist common attacks, such as statistical, differential, known plaintext attack and chosen plaintext attack. Efficient analysis indicates that it has low computation and time complexity. Therefore, it has excellent application prospect.
\par 
This paper is arranged as follows: Section~\ref{section:section2} describes the dynamics of the 2D-Duffing map. In Section\ref{section:section3}, we introduce a chaotification approach to enhance chaos complexity of the 2D-Duffing map, and then calculate the performance of the enhance map. Section~\ref{section:section4} introduces an  image encryption algorithm based on the enhanced Duffing map. Simulation results for  image encryption using some existing schemes and our proposed scheme are presented in  Section \ref{section:section5}. Section \ref{section:section6} is left to perform the security analysis of encrypted images.  Finally, Section \ref{section:section7} concludes the results.
\section{The Duffing model}
\label{section:section2}
The Duffing map, also known as the Holmes map \cite{ding1991time}, is a 2D discrete-time chaotic system, given by,
\begin{flalign}
\label{1}
\begin{cases}
\begin{aligned}
x_{1}(n+1)&=x_{2}(n),\\
x_{2}(n+1)&=-\beta x_{1}(n)+\alpha x_{2}(n)-x_{2}^{3}(n),
\end{aligned}
\end{cases}
\end{flalign}
where the parameters $\alpha$ and $ \beta$  are  positive. 

\subsection{Stability of equilibrium points}\label{sec-eqbm}
From a graphical point of view, a point $ E $ is said to be an equilibrium point of the function $ G(x) $ only if $ G^{n}(E)=E $. Thus, one can obtain  the equilibrium points of the system \eqref{1} by reducing its dimension as follows: 
\begin{flalign}
\label{2}
x_{1}^{(v)}&=-\beta x_{1}^{(v)}+\alpha x_{1}^{(v)}-(x_{1}^{(v)})^{3},
\end{flalign}
where $v= 1, 2, \dots$ For the parameters $ \alpha=2.75 $ and $ \beta=0.2 $, the equilibrium points can be obtained as:
\begin{flalign*}
\begin{cases}
\begin{aligned}
&E_{1}=(0, 0),\nonumber\\
&E_{2}=(1.244, 1.244),\nonumber\\
&E_{3}=(-1.244, -1.244).
\end{aligned}
\end{cases}
\end{flalign*}

\begin{figure}[t]
	\centering
	\includegraphics[width=8cm, height=5cm]{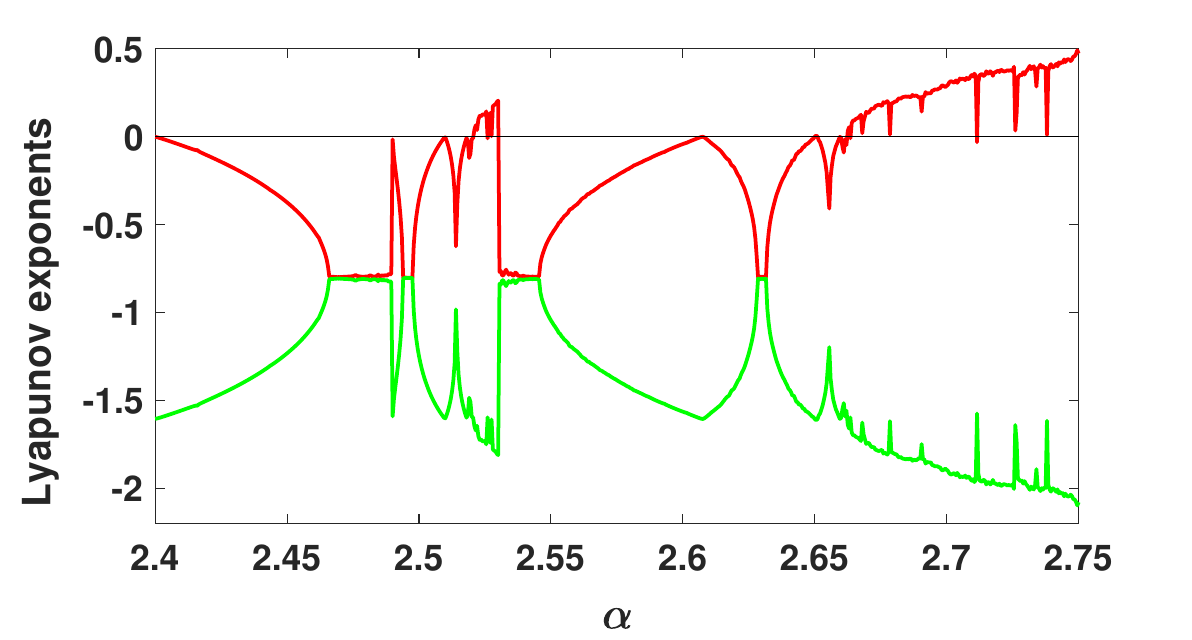}
	\centering
	\caption{Lyapunov exponents of the Duffing map \eqref{1} with the parameter $ \beta=0.2 $  and   the initial condition $ (0.25, 0.77) $.}
	\label{fig:lyapunov1}
\end{figure}

\begin{figure}[!bh]
	\centering
	\includegraphics[width=8cm, height=5cm]{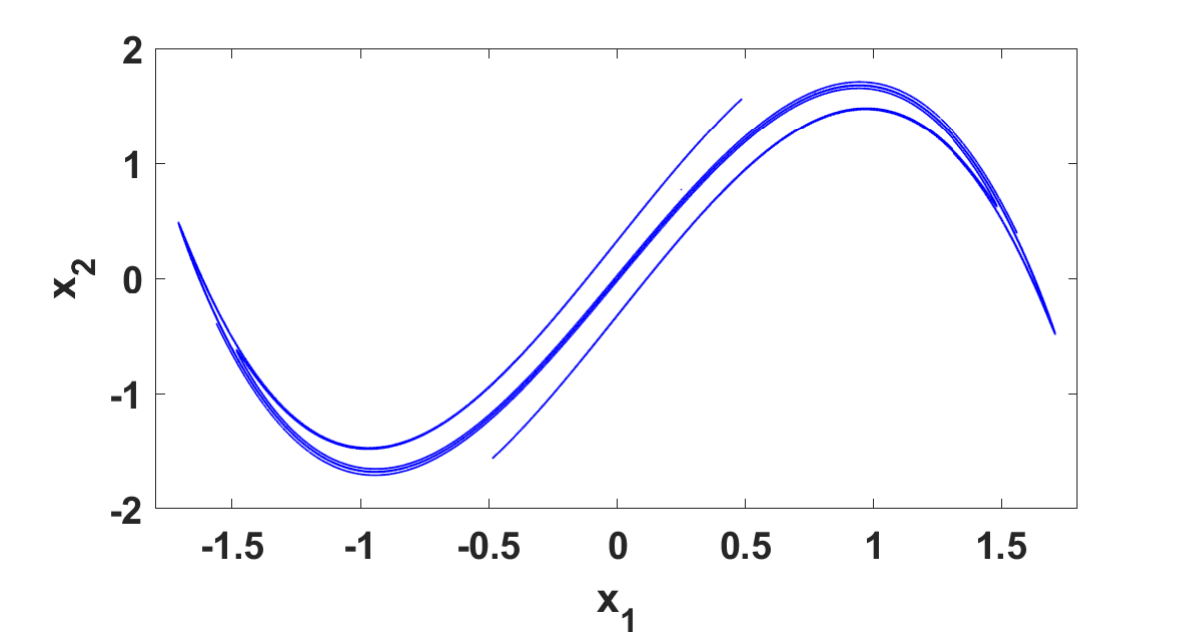}
	\centering
	\caption{Chaotic attractor of the Duffing map \eqref{1} with the parameter $ \beta=0.2 $, $\alpha=2.75 $ and for the initial conditions $ (0.25, 0.77) $.}
	\label{fig:phase2}
\end{figure}
The stability of the   equilibrium points is determined by the following Jacobian matrix of the Duffing map \eqref{1}.
\begin{flalign*}
\begin{aligned}
J
=\begin{pmatrix}
\frac{\partial f_{1}}{\partial x_{1}} &\ \  \frac{\partial f_{1}}{\partial x_{2}} \\
\\
\frac{\partial f_{2}}{\partial x_{1}} &\ \  \frac{\partial f_{2}}{\partial x_{2}} 
\end{pmatrix}.
\end{aligned}
\end{flalign*}
Next, the Duffing map \eqref{1} can be linearized with respect to an equilibrium point $ E_{i}=(x_{1}^{\ast}, x_{2}^{\ast}) $ by
\begin{flalign*}
\begin{aligned}
J_{E_{i}}=\begin{pmatrix}
0& \ \ 1\\
-\beta &\ \  \alpha-3(x_{2}^{\ast})^{2}\\
\end{pmatrix}.
\end{aligned}
\end{flalign*}
The corresponding eigenvalues at   $ E_{i} $ can be obtained by solving  
$\det(\lambda I-J_{E_{i}})=0$, which  yields 
\begin{flalign*}
\lambda^{2}+\left(3(x_{2}^{\ast})^{2}-\alpha\right) \lambda+\beta=0.
\end{flalign*}
Thus, the eigenvalues are given by
\begin{flalign*}
\begin{cases}
\begin{aligned}
\lambda_{1}=&\frac{3(x_{2}^{\ast})^{2}-\alpha-\sqrt{\left(3(x_{2}^{\ast})^{2}-\alpha\right)^{2}-4\beta}}{2} \\
\lambda_{2}=&\frac{-3(x_{2}^{\ast})^{2}+\alpha+\sqrt{\left(3(x_{2}^{\ast})^{2}-\alpha\right)^{2}-4\beta}}{2}
\end{aligned}
\end{cases}.
\end{flalign*}
In   discrete dynamical systems, the stability of equilibrium points is dependent on the corresponding eigenvalues. If an eigenvalue lies the interval $ [-1, 1] $, then the equilibrium point is said to exhibit  a stable state. Otherwise, it represents an unstable state. For the parameters $ \alpha=2.75 $ and $ \beta=0.2 $, the stability of the  equilibrium points of the Duffing map~\eqref{1} is illustrated in Table~\ref{tb:tabel1}. Clearly, the Duffing map has one unstable   and two stable equilibrium points.

\begin{figure}[b]
	\centering
	\begin{subfigure}[h]{0.42\textwidth}
		\includegraphics[width=8cm, height=4.5cm]{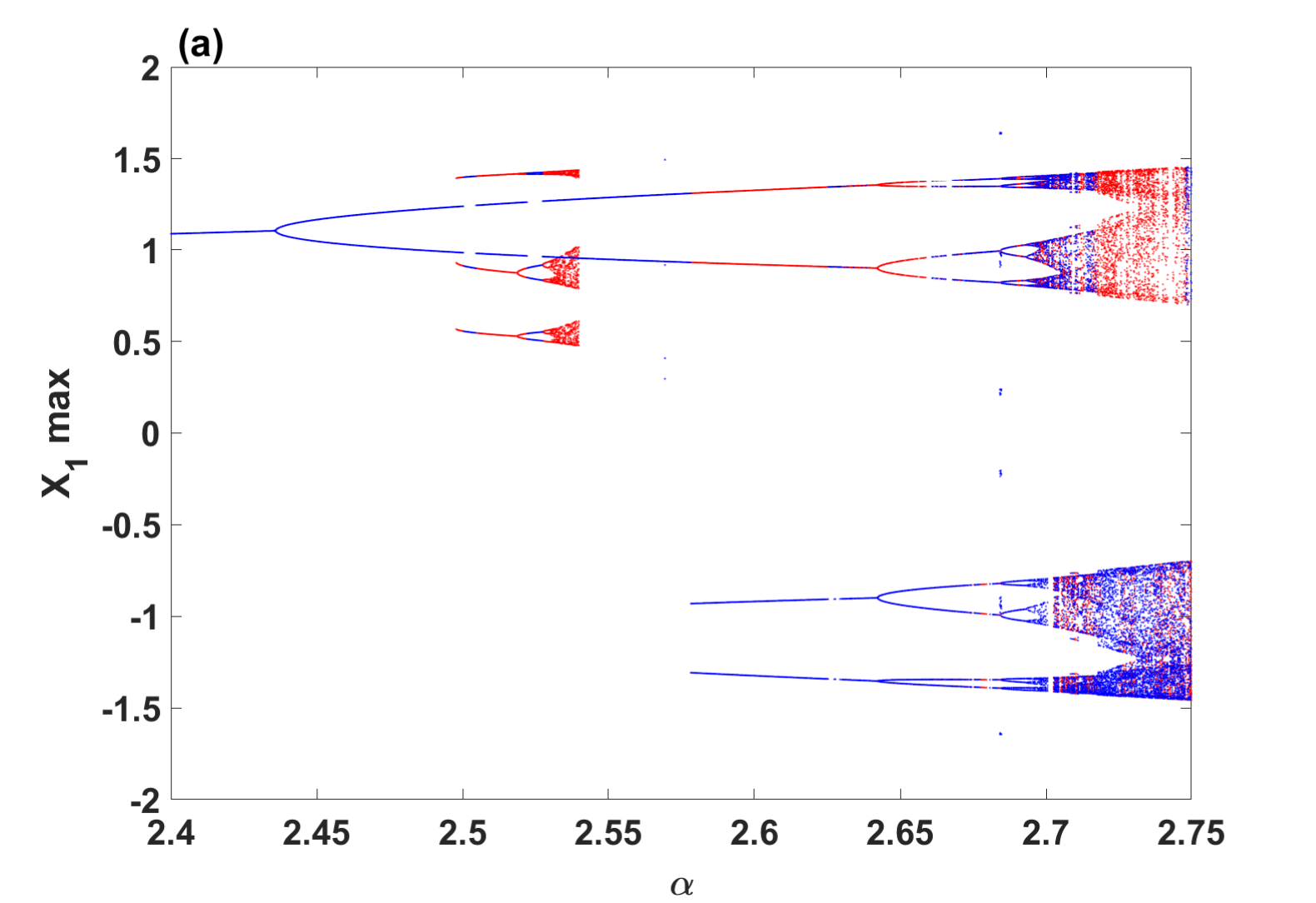}
	\end{subfigure}
	
	\begin{subfigure}[h]{0.42\textwidth}
		\includegraphics[width=8cm, height=4.5cm]{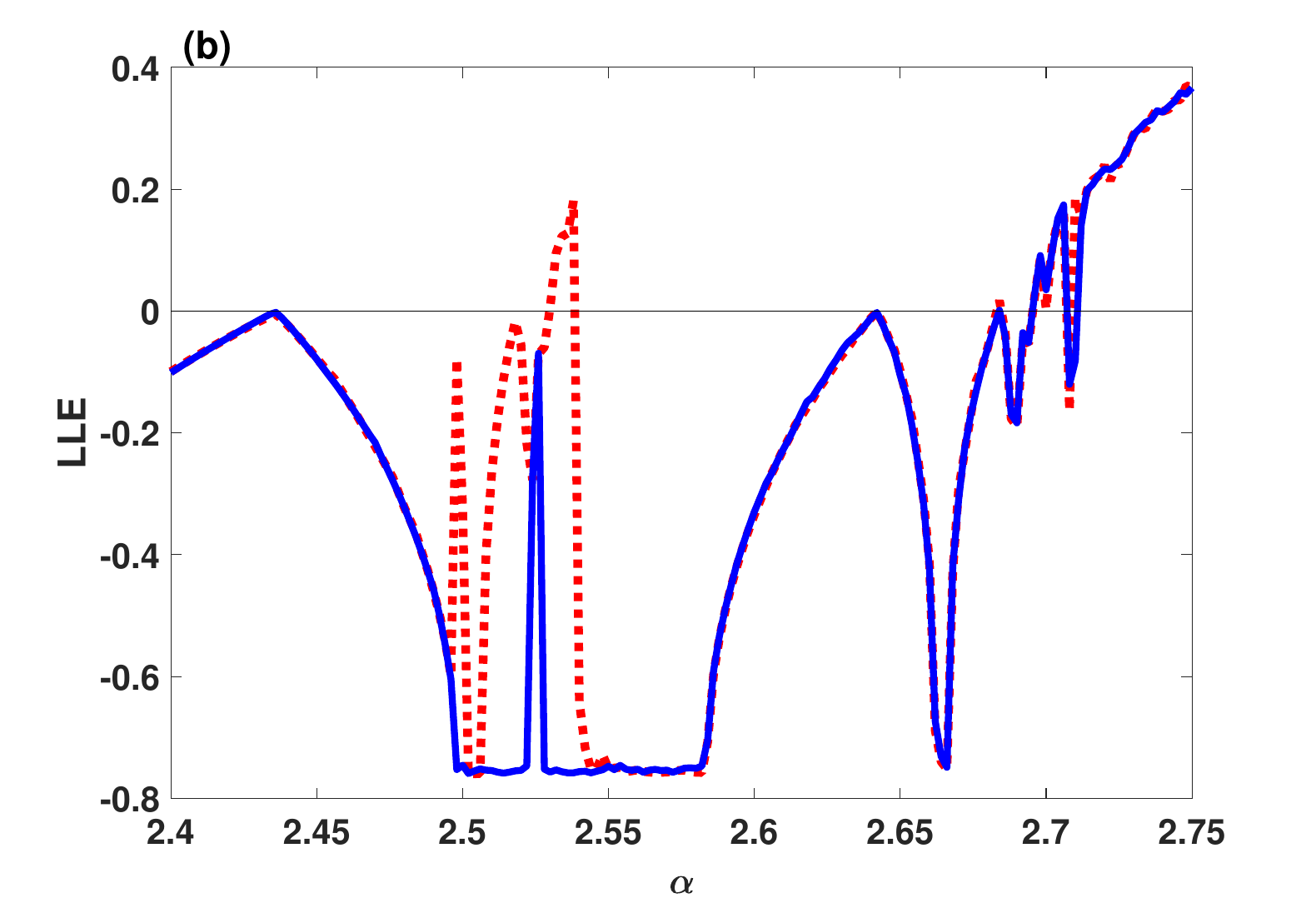}
	\end{subfigure}
	\caption{Coexisting two attractors with the parameter $ \beta=0.218 $, and under two sets of initial states in which the blue and red orbits are initiated from $ (0.25, 0.97) $ and $ (0.25, 0.77) $ respectively: (a) Bifurcation diagram with respect to the variable $ x_{1} $; (b)	Largest Lyapunov exponents.}
	\label{fig:LLE}
\end{figure}

\begin{table}[h]
	\setlength{\tabcolsep}{8pt}
	\renewcommand{\arraystretch}{1.5}	
	\captionof{table}{The equilibria of Duffing map~\eqref{1} and their stability for the parameters $ \beta=0.2 $ and $ \alpha=2.75 $.}
	\label{tb:tabel1}
	\begin{tabular}{c c c c} 
		\hline
		
		\hline
		Equilibria & $ \lambda_{1} $& $ \lambda_{2}$ &Stability analysis\\
		\hline
		
		\hline
		
		$ E_{1} $ &$ -2.6752 $&$ 2.6752$ &unstable equilibrium \\ 
		
		$ E_{2} $ &$ 0.1123 $&$ -0.1123$ &stable equilibrium \\ 
		
		$ E_{3} $ &$ 0.1123 $&$ -0.1123$ &stable equilibrium \\ 
		\hline
		
		\hline
	\end{tabular}
\end{table}

\subsection{Self-excited chaotic attractor}
To investigate the dynamical behaviors of the map~\eqref{1}, we calculate the Lyapunov exponents (LE) as depicted in Figure~\ref{fig:lyapunov1} corresponding to the initial conditions $ (0.25, 0.77) $. It is seen that the map~\eqref{1} exhibits two different behaviors in one of which it exhibits chaos where the largest Lyapunov exponent (LLE) is positive, and in the other it shows periodic behaviors with LLE is zero or negative. Figure~\ref{fig:phase2} demonstrates the chaotic behaviors of the Duffing map corresponding to the parameters  $ \alpha=2.75 $ and $ \beta=0.2 $. Furthermore, since the map~\eqref{1} has one unstable equilibrium point corresponding to a different set of   parameters $ \alpha=2.75 $ and $ \beta=0.2 $,  the chaotic attractor in Figure~\ref{fig:phase2} is self-excited \cite{natiq2018self}.  

\begin{figure}[t]
	\centering
	\begin{subfigure}[h]{0.22\textwidth}
		\includegraphics[width=4.3cm, height=4cm]{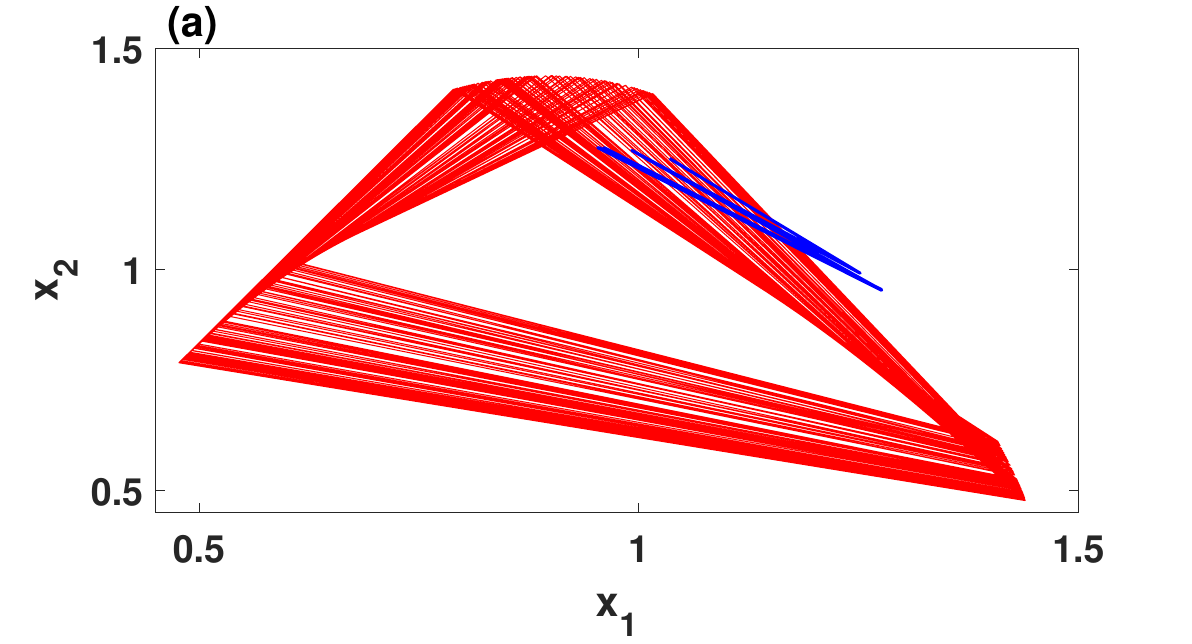}
	\end{subfigure}
	\begin{subfigure}[h]{0.22\textwidth}
		\includegraphics[width=4.3cm, height=4cm]{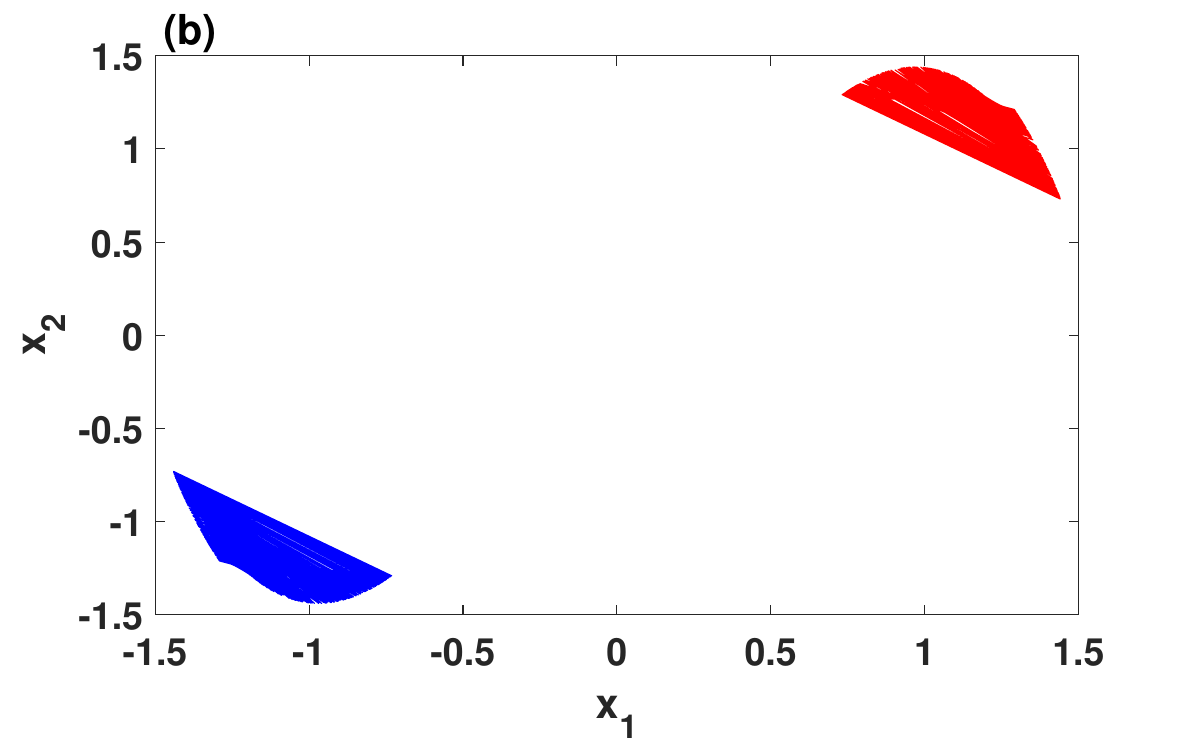}
	\end{subfigure}
	
	\caption{Multistability behaviors of the Duffing map~\eqref{1} under different initial conditions $ (0.25, 0.77) $ (red) and $ (0.25, 0.97) $ (blue): (a) the coexistence of chaotic and periodic orbit with the parameters $ \alpha=2.5392 $, and $ \beta=0.218 $; (b) the coexistence of two chaotic attractors with the parameters $ \alpha=2.73 $, and $ \beta=0.218 $.}
	\label{fig:phase3}
\end{figure}
\subsection{Coexisting attractors}
Multistability behaviors or coexisting attractors indicate that the nonlinear dynamical system  can produce two or more attractors by changing the initial conditions. The coexistence of attractors is a phenomenon that happens in nonlinear dynamical systems due to the high sensitivity of such systems to their initial conditions. In this subsection, we demonstrate the existing of  multistability behaviors in the Duffing map \eqref{1} by considering an appropriate set of initial conditions. Such complicated behaviors of the Duffing map have not been studied in the previous studies. 
\par 
For a fixed value of $\beta$, i.e., $ \beta=0.218 $, when the parameter $ \alpha $ is varied from $ 2.4 $ to $ 2.75 $, the coexisting bifurcation model and the LE of the Duffing map \eqref{1} corresponding to the initial conditions $ (0.25, 0.97) $ (blue), $ (0.25, 0.77) $ (red) are plotted   as shown in Figure~\ref{fig:LLE}. It is seen that   the coexistence of self-excited chaotic attractors and the periodic orbits occur mainly in the region   $ 2.532\leq \alpha \leq 2.539 $. However, two self-excited chaotic attractors can also coexist in some other region of $ 2.71\leq \alpha \leq 2.74 $. To further visualize these interesting features, we consider two other values of  $ \alpha $, i.e.,  $\alpha= 2.5392 $ and $ 2.73 $. The results are displayed in  Figure~\ref{fig:phase3} (a) and (b)  respectively. As can be seen in Fig.~\ref{fig:phase3} (a), the Duffing map \eqref{1} shows the coexistence of a chaotic attractor and periodic orbit. Furthermore, this map shows in Fig.~\ref{fig:phase3} (b) the coexistence of two chaotic attractors located in different positions depending on the location of the two unstable equilibria $ E_2 $ and $ E_3 $.

It is to be noted that  the keys in the chaos-based cryptosystems are typically formed by means of the initial conditions and the parameters of the chaotic map. However, when the chaotic map  exhibits multistability behaviors, such as the coexistence of chaotic attractors and periodic orbits, the corresponding cryptosystems will be insecure. In this situation,  one requires  to introduce an efficient chaotification technique on 2D Duffing map~\eqref{1} for enhancing chaos in the non-chaotic regions.    

\begin{figure}[t]
	\centering
	\includegraphics[width=9cm, height=5cm]{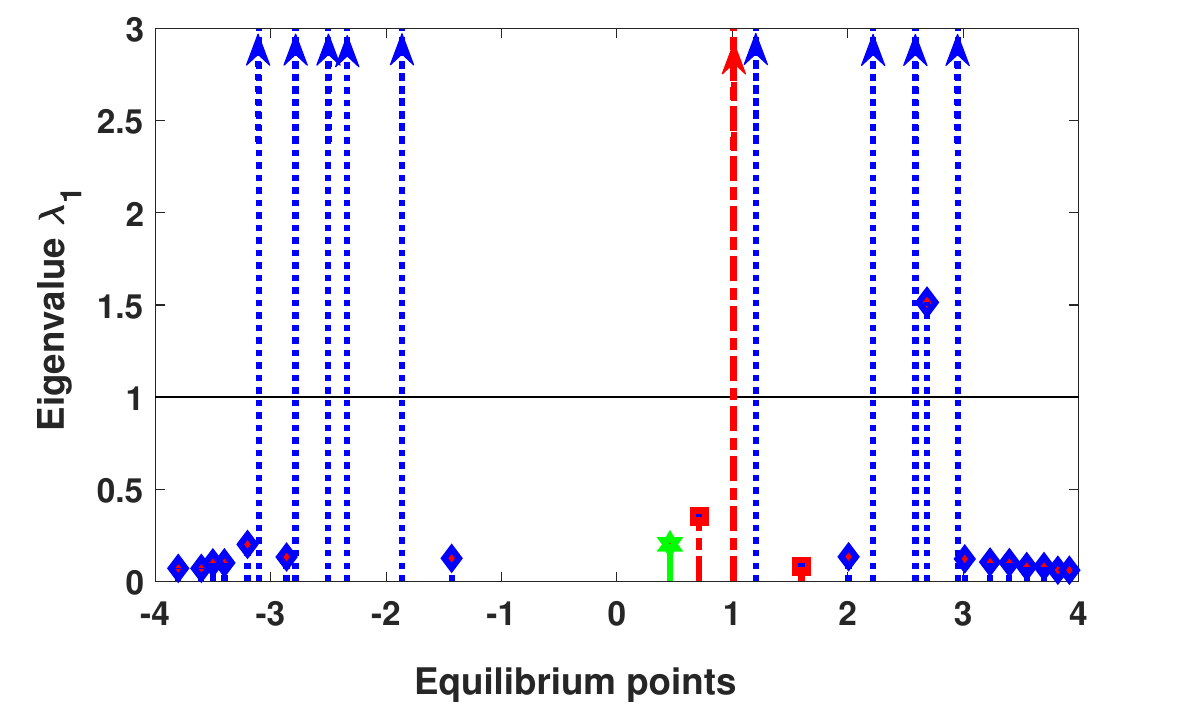}
	\centering
	\caption{Equilibrium points of the enhanced Duffing map \eqref{4} and its stability when the parameters $ \alpha=2.75 $ and $ \beta=0.2 $: 1) for $ A=1 $ and $ B=1 $, there is only one stable equilibrium point (green color); 2)  for $ A=2 $ and $ B=3 $, there are three equilibria (red color) in which two stable and one unstable; 3) for $ A=15 $ and $ B=3.7 $, there are 25 equilibria (blue color) in which 15 stable and 10 unstable.}
	\label{fig:fixed2}
\end{figure}

\section{Sine-Cosine chaotification technique}
\label{section:section3}
This section proposes a new chaotification technique that uses two trigonometric functions as nonlinear transforms to the outputs of the Duffing map \eqref{1}. The Sine and Cosine functions are applied to enhance the chaos and complexity of the Duffing   map in the chaotic region. Moreover, these functions can also be used to produce chaos in the non-chaotic regions. 

\begin{figure}[t]
	\centering
	\includegraphics[width=9cm, height=4.5cm]{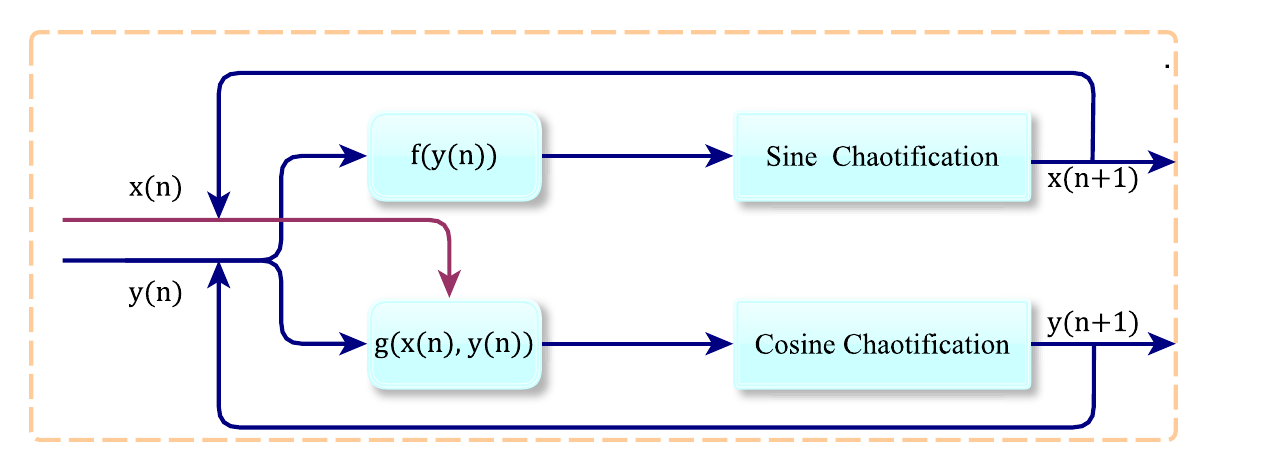}
	\centering
	\caption{Structure of the enhanced Duffing map.}
	\label{fig:diagram1}
\end{figure}

The structure of the proposed technique is shown in Figure~\ref{fig:diagram1}, where $  f\left( y(n)\right)  $ and $g \left( x(n), y(n) \right) $ are two seed maps taken from  Eq. \eqref{1}.   Mathematically, the proposed map can be defined as follows
\begin{flalign}
\label{4}
\begin{cases}
\begin{aligned}
x(n+1)&=A\sin(y(n)),\\
y(n+1)&=B\cos(-\beta x(n)+\alpha y(n)-y^{3}(n)),
\end{aligned}
\end{cases}
\end{flalign}
where $ A ~(>0)$, and $ B~(>0) $ are  parameters.

\subsection{Stability analysis of the enhanced Duffing map}
In the previous section \ref{sec-eqbm}, we have seen that   the Duffing map \eqref{1} has three equilibrium points in which two of them are stable and the other unstable, as illustrated in Table~\ref{tb:tabel1}. So, it is interesting to know whether the enhanced Duffing map \eqref{4} generates the same numbers or different numbers of equilibria when $ \alpha=2.75 $ and $ \beta=0.2 $. Besides, it is important to investigate the stability of these equilibria.

We note that the eigenvalues of the enhanced Duffing map satisfy the relation $ \lambda_1=-\lambda_2 $.  Three cases may be of interest to investigate  the stability of equilibrium points of the enhanced Duffing map: 1) when $ A=1 $ and $ B=1 $, the enhanced Duffing map \eqref{4} has only one stable equilibrium point, as shown in the green color of Figure~\ref{fig:fixed2}; 2) when $ A=2 $ and $ B=3 $, the enhanced Duffing map \eqref{4} has three different equilibria in which two of them are stable and one unstable, as shown in the red color of Figure~\ref{fig:fixed2}. 3) when $ A=15 $ and $ B=3.7 $, the enhanced Duffing map \eqref{4} has 25 different equilibria in which 15 equilibria are stable and 10 equilibria are unstable, as shown in the blue color of Figure~\ref{fig:fixed2}.

Thus, it can be concluded that depending on the values of the parameters $ A $ and $ B $, the enhanced Duffing map~\eqref{4} can generate equilibrium points less than or equal to  or greater than those of the  Duffing map~\eqref{1}. 

\begin{figure*}
	\centering
	\begin{minipage}[b]{0.48\textwidth}
		\begin{subfigure}[b]{0.49\linewidth}
			\includegraphics[width=9cm, height=2.7cm]{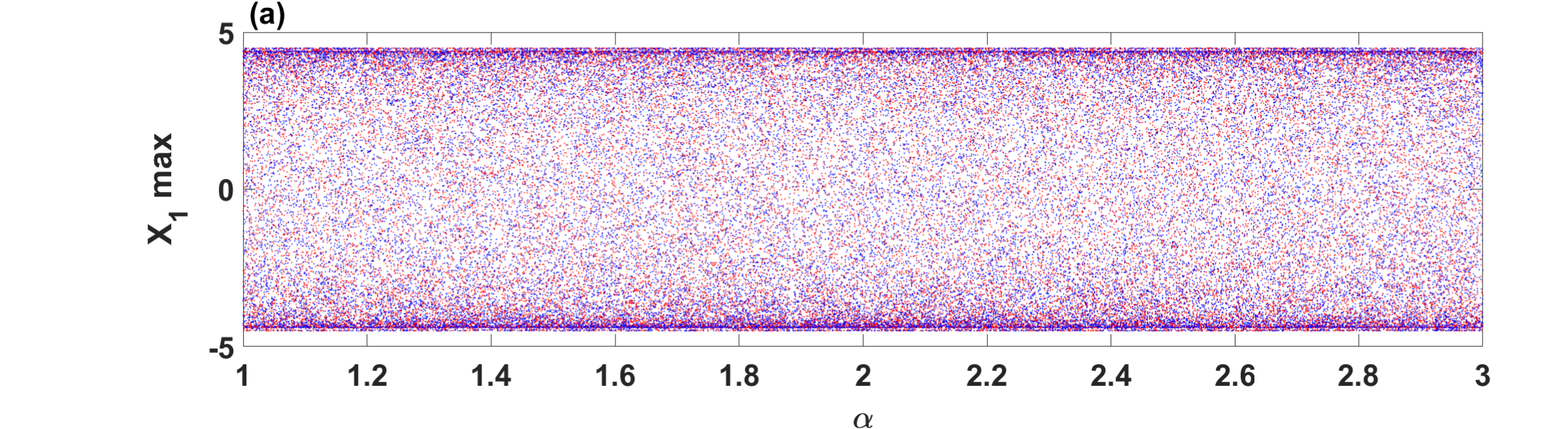}
		\end{subfigure}
		
		\begin{subfigure}[b]{0.49\linewidth}
			\includegraphics[width=9cm, height=2.7cm]{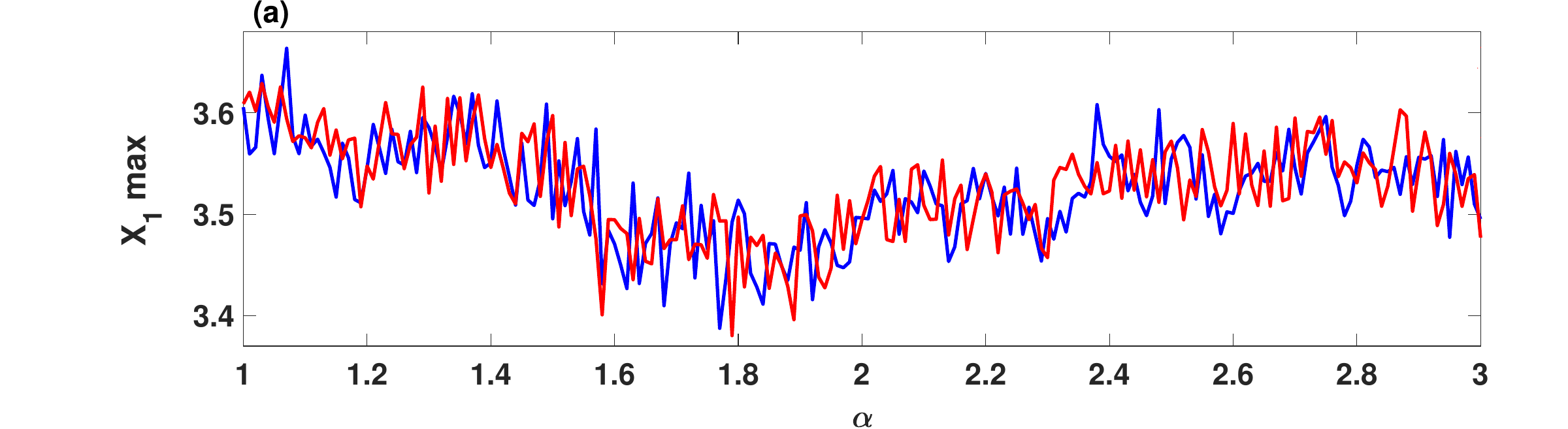}
		\end{subfigure}
	\end{minipage}
	\hfill
	\begin{subfigure}[b]{0.48\textwidth}
		\hspace{-0.4cm}
		\vspace{-0.25cm}
		\includegraphics[width=8cm, height=5.5cm]{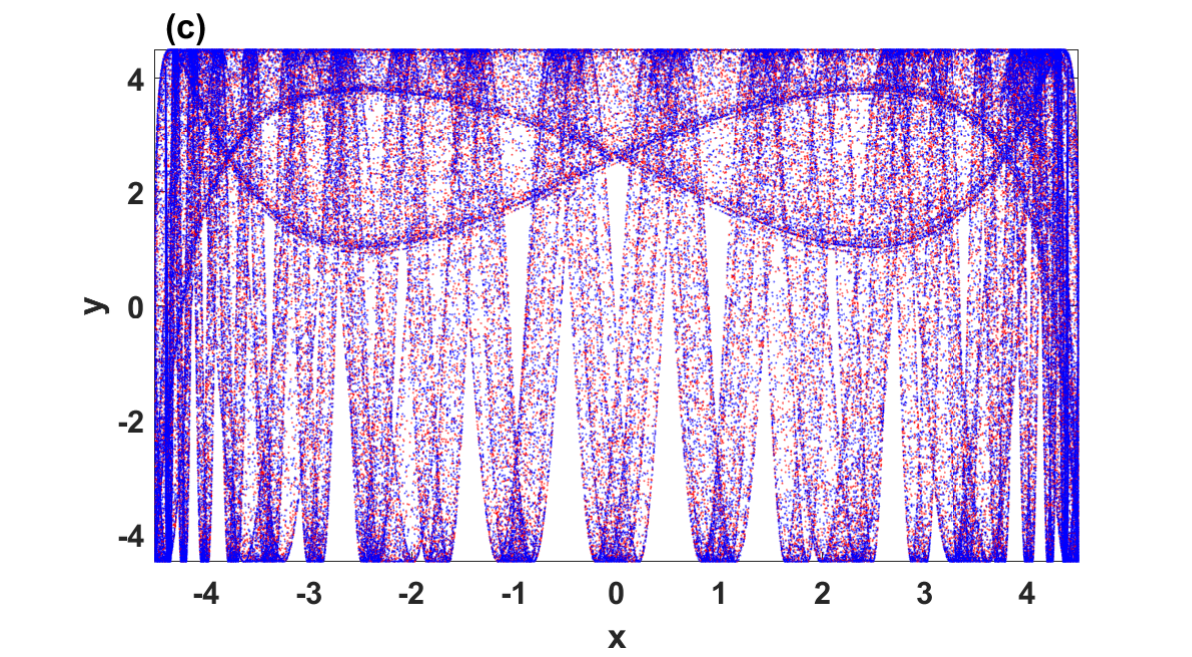}
	\end{subfigure}
	\vspace{0.25cm}
	\caption{Dynamics of the enhanced Duffing map \eqref{4} under the initial conditions $ (0.25, 0.77) $ (red) and $ (0.25, 0.97) $ (blue) with the parameters $ \beta=0.218 $, $ A=B=4.5 $: (a) the coexisting bifurcation model; (b) Largest Lyapunov exponents; (c) phase space with $ \alpha=1 $.}
	\label{fig:bifurcation2}
\end{figure*}

\begin{figure}[!bh]
	\centering
	\begin{subfigure}[h]{0.5\textwidth}
		\includegraphics[width=8cm, height=4cm]{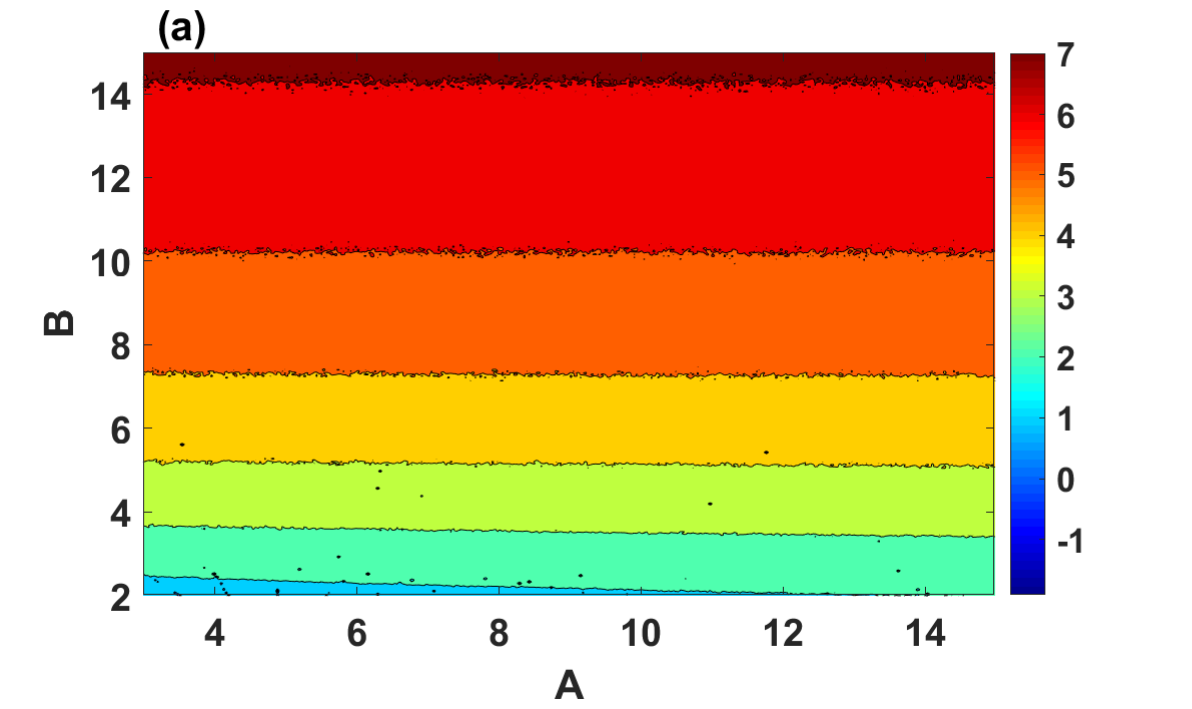}
	\end{subfigure}
	
	\begin{subfigure}[h]{0.5\textwidth}
		\includegraphics[width=8cm, height=4cm]{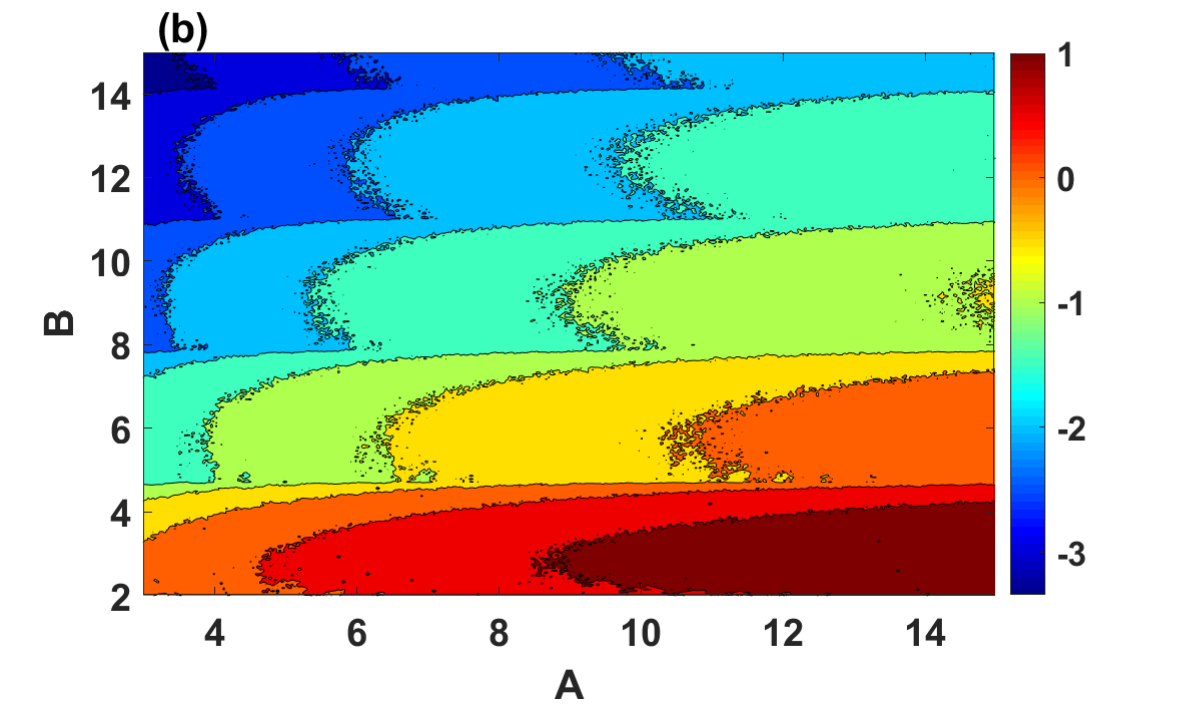}
	\end{subfigure}
	\caption{Hyperchaotic diagram of the enhanced Duffing map~\eqref{4} based on Lyapunov exponents with $ \beta=5 $, $\alpha=1$: (a) the largest Lyapunov exponent; (b) the lowest Lyapunov exponent.}
	\label{fig:contour1}
\end{figure}

\subsection{Enhancing chaos of Duffing map}
Since the number of equilibria of the enhanced map \eqref{4} can be larger than those of the original map when the parameters $ A $ and $ B $ are large enough [See the third case in Figure~\ref{fig:fixed2}], it is quite reasonable to assume that increasing the equilibria of a dynamical system in a limited range can enhance chaos with overlapping coexisting attractors.

In order to graphically demonstrate the above features, we consider the parameters of the enhanced map \eqref{4} as $ \beta=0.218 $, $ A=B=4.5 $ for which   the map has $37$ equilibrium points distributed within the range $ [-3.89, 3.97] $. Figure~\ref{fig:bifurcation2} (a) and (b) depict the coexisting bifurcation model and LLE of the Duffing map \eqref{1} corresponding to two sets of  initial conditions $ (0.25, 97) $ (blue), $ (0.25, 0.77) $ (red). Clearly, the chaotic regions of the 2D Duffing map are enhanced and the non-chaotic regions shift to the chaotic regions. Furthermore, the two coexisting chaotic attractors of the enhanced map~\eqref{4} are overlapped, and occupied a much larger region in the 2D phase space, as can be seen in Figure~\ref{fig:bifurcation2} (c). 
\subsection{Generating hyperchaotic behaviors}
A nonlinear dynamical system exhibits hyperchaotic behaviors only when it has at least two positive values of the LEs. So,    the 2D Duffing map~\eqref{1} exhibits no hyperchaotic behavior as it has only one positive value of the LE with some parameter values (See Figure~\ref{fig:lyapunov1}). Typically, the trajectory of a dynamical system with hyperchaotic behavior is more difficult to  predict than chaotic ones, and so   is  desirable for cryptography applications.   

To examine the occurrence of hyperchaotic behaviors in the enhanced Duffing map~\eqref{4}, we  calculate the LEs when both the  parameters $ A $ and $ B $ vary  as shown in Figure~\ref{fig:contour1}. The LLE of the enhanced map~\eqref{4} is illustrated in Figure~\ref{fig:contour1} (a), while the lowest LE is illustrated in Figure~\ref{fig:contour1} (b). It is seen  that the enhanced map shows chaotic behaviors with   higher values of the LLE when $ B\in(7.3, 15] $ and for any value  of the parameter $ A $. However, the hyperchaotic behaviors of the enhanced map are  observed in two regions as follows: 1) $ A\in[12, 15] $ and $ B\in[4, 7.3)$; 2) $ A\in[3, 15] $ and $ B\in[2, 4] $.

\begin{figure*}
	\begin{center}
		\begin{subfigure}[h]{0.3\textwidth}
			\includegraphics[width=5.5cm, height=4cm]{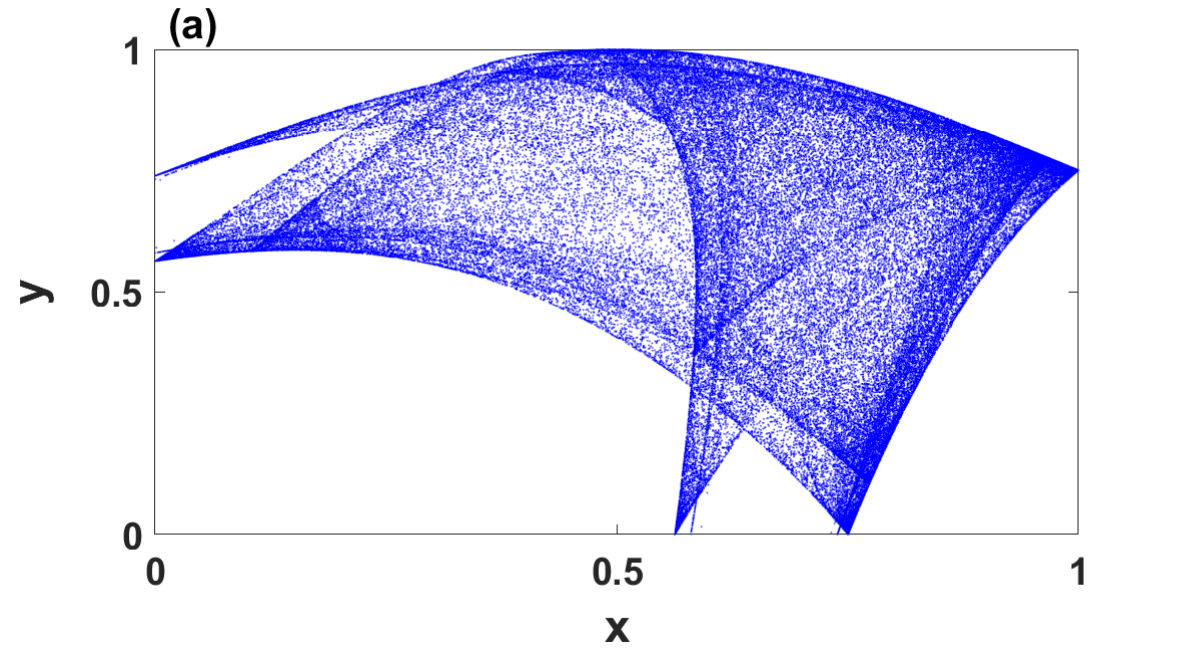}
		\end{subfigure}
		\begin{subfigure}[h]{0.3\textwidth}
			\includegraphics[width=5.5cm, height=4cm]{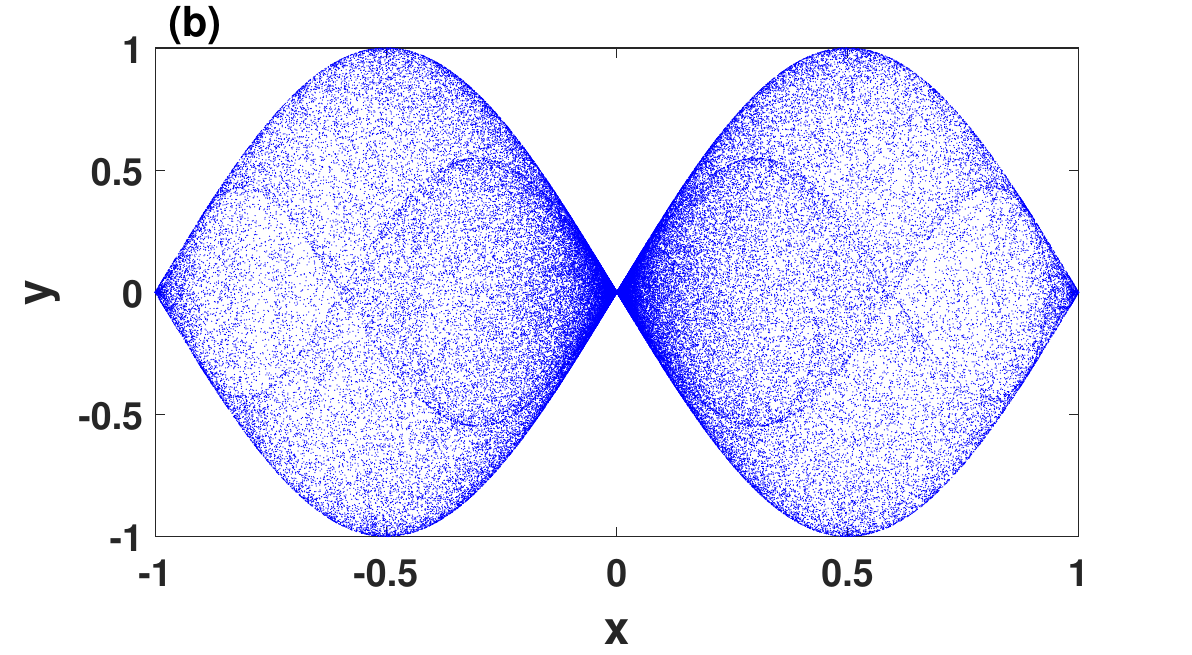}
		\end{subfigure}
		\begin{subfigure}[h]{0.3\textwidth}
			\includegraphics[width=5.5cm, height=4cm]{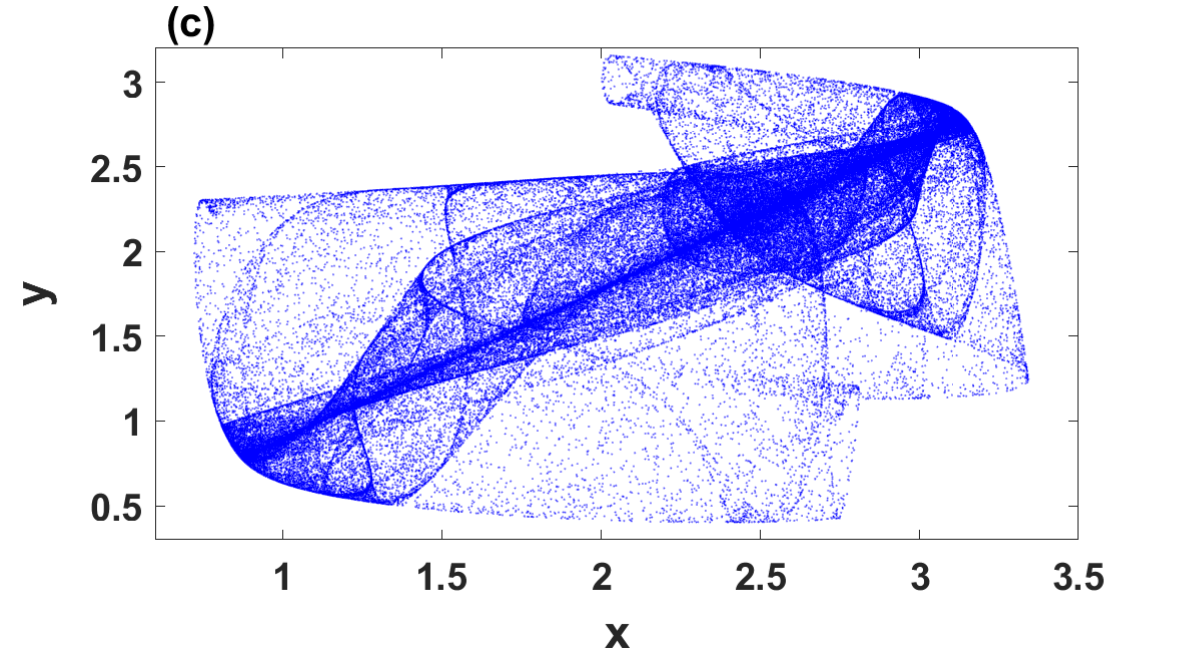}
		\end{subfigure}
		
		\begin{subfigure}[h]{0.3\textwidth}
			\includegraphics[width=5.5cm, height=4cm]{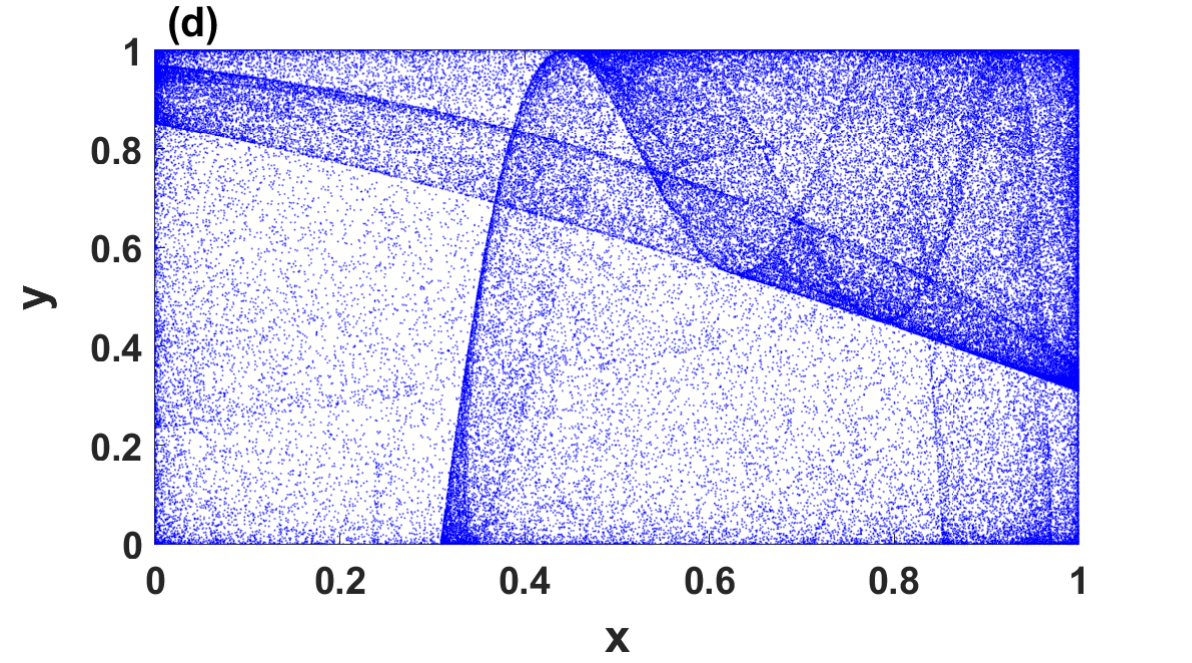}
		\end{subfigure}
		\begin{subfigure}[h]{0.3\textwidth}
			\includegraphics[width=5.5cm, height=4cm]{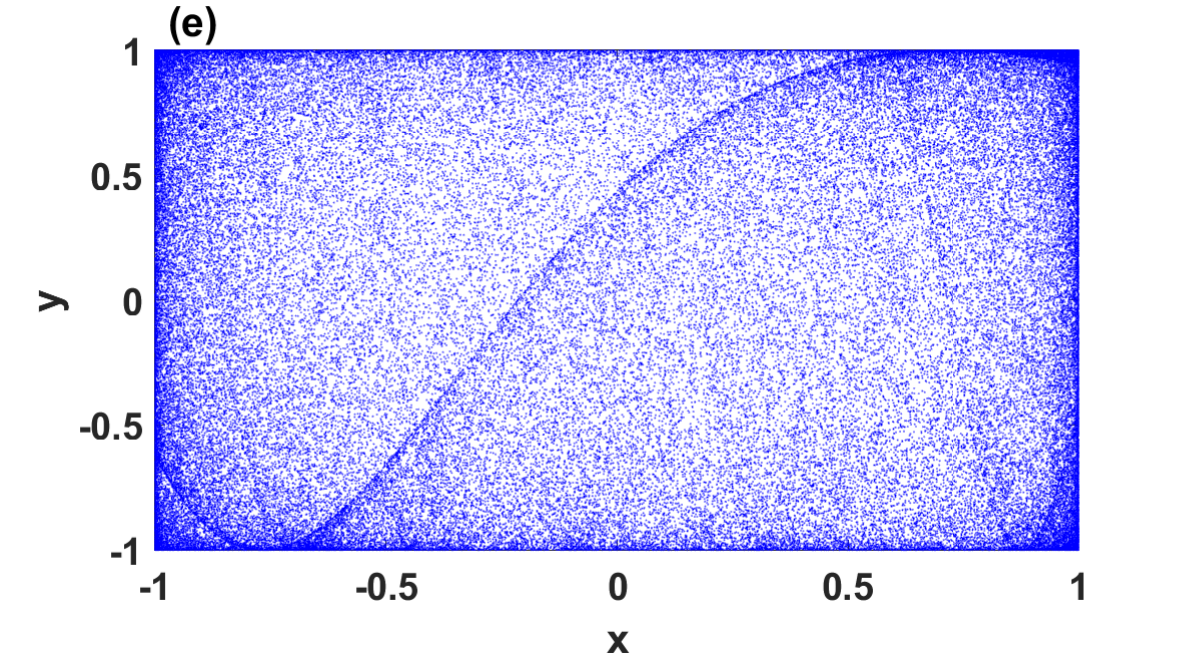}
		\end{subfigure}
		\begin{subfigure}[h]{0.3\textwidth}
			\includegraphics[width=5.5cm, height=4cm]{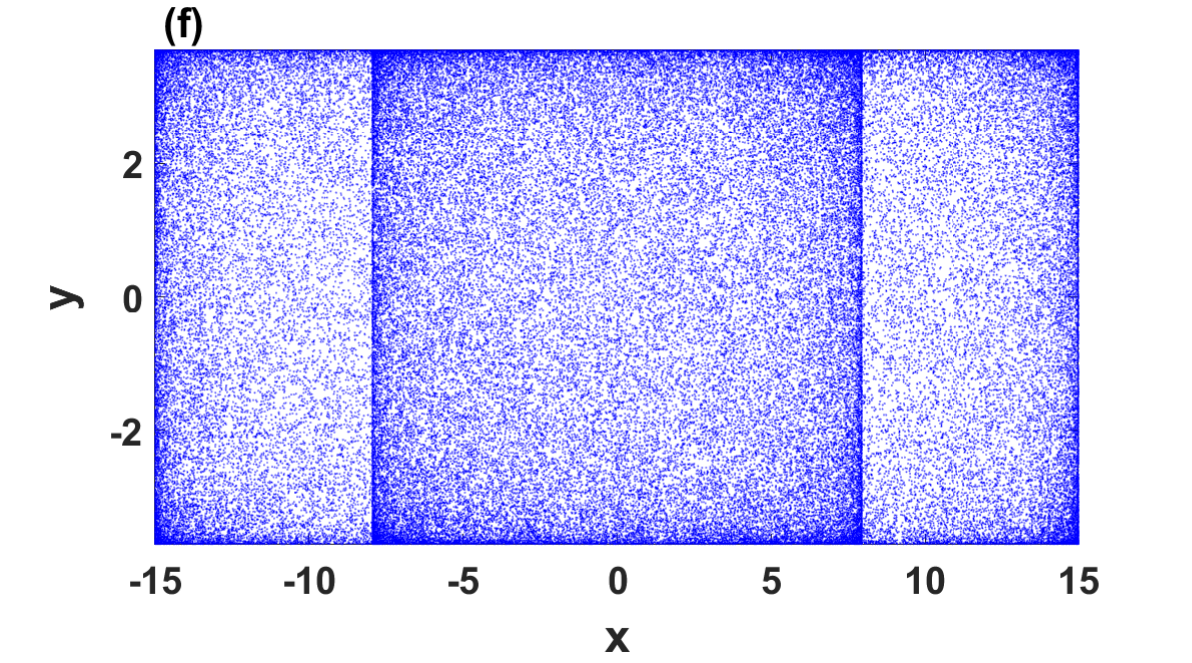}
		\end{subfigure}
	\end{center}
	\caption{Chaotic and hyperchaotic attractors of different 2D maps: (a) 2D-SLMM~\cite{hua20152d}; (b) 2D-SIMM  \cite{liu2016fast}; (c) 2D Ushiki map \cite{gao2009study}; (d) 2D-LASM \cite{hua2016image}; (e) 2D-LICM \cite{cao2018novel}; (f) the enhanced Duffing map~\eqref{4}.}
	\label{fig:attractor}
\end{figure*}

In order to verify the features of the enhanced Duffing map~\eqref{4} that the hyperchaotic attractor  has high level of complexity  and spreads in a wide region of the 2D phase space, we plot the hyperchaotic attractor of the enhanced map as well as the attractors of other chaotic and hyperchaotic maps as shown in Figure~\ref{fig:attractor}.  Here, the 2D-SLMM~\cite{hua20152d}, 2D-SIMM \cite{liu2016fast}, 2D-LASM \cite{hua2016image}, and 2D-LICM \cite{cao2018novel} are considered as hyperchaotic maps, while the 2D Ushiki \cite{gao2009study}  as a chaotic map.

Figure~\ref{fig:attractor} shows that the attractor of the enhanced map with the parameters $ \alpha=1 $, $ \beta=5 $, $ A=15 $, $ B=3.7 $ occupies the whole 2D phase space with regions $ x\in[-15, 15] $ and $ y\in[-3.7, 3.7] $. This means that the enhanced map generates some extreme unpredictable sequences, and its ergodicity property is much better than other maps.

\begin{figure}[!bh]
	\begin{center}
		\includegraphics[width=8.5cm, height=4.5cm]{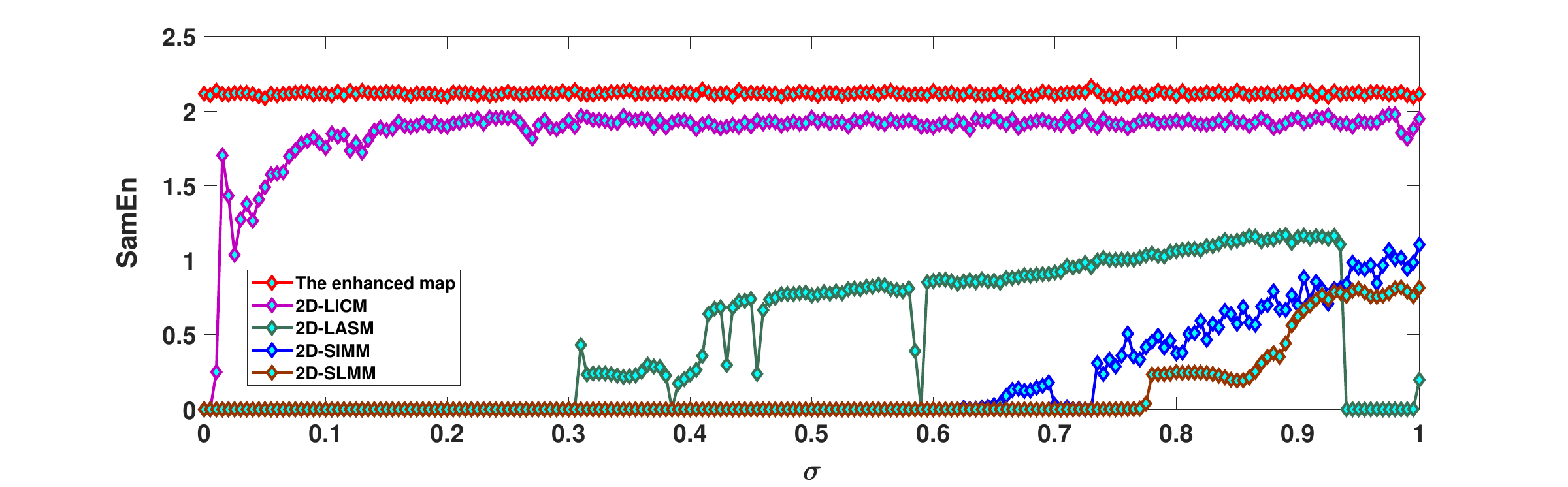}
	\end{center}
	\caption{SamEn results of different chaotic and hyperchaotic maps, where parameter $ \sigma $ represents $ \alpha, a, a_{1}, a_{2}, \alpha_{1} $, for the enhanced map \eqref{4}, 2D-LICM \cite{cao2018novel}, 2D-LASM \cite{hua2016image}, 2D-SIMM \cite{liu2016fast}, and 2D-SLMM \cite{hua20152d}, respectively.}
	\label{fig:complexity1}
\end{figure}

\subsection{Complexity based Sample Entropy}
Richman et al. \cite{richman2000physiological} introduced an approach to develop an Approximate Entropy, which is widely used as a measure for estimating the time series complexity. Several analyses have demonstrated that the developed measure, namely, the Sample Entropy (SamEn)  is more  accurate than Approximate Entropy.

To illustrate the complexity of the enhanced map~\eqref{4}, Figure~\ref{fig:complexity1} depicts the SamEn results of the enhanced map and different chaotic and hyperchaotic maps. The parameters of the enhanced map \eqref{4} for this figure are considered as $ \beta=5 $, $ A=15 $, and $ B=3.7 $. Clearly, the enhanced map~\eqref{4} has the largest SamEn values implying that one needs more information to predict the generated sequences by this map.

\subsection{Randomness analysis}
The randomness of  two  hyperchaotic sequences $ \{x\} $ and $ \{y \}$, so generated by the enhanced Duffing map~\eqref{4}, can be examined by several randomness evaluation methods. Here, we use the software package of FIPS 140-2, which mainly consists of three different tests. For each test, the p- value is derived to reflect the randomness level. A chaotic sequence can pass the test when the derived p-value is within a range of $ [10^{-4}, 1-10^{-4}] $. The experimental results are shown in Table~\ref{tab3}. As can be seen that the   sequences $ \{x\} $ and $ \{y \}$ pass all the statistical tests, which means that these two sequences are reliable PRNG, and have excellent randomness property.

In summary, Lyapunov exponents, trajectory, SamEn, and FIPS 140-2 have demonstrated that the enhanced map~\eqref{4} exhibits decent ergodicity property, wide hyperchaotic behavior, high level of complexity and randomness. As a result, the enhanced map would be very promising for cryptography applications.

\begin{table*}
	\caption{Randomness analysis test with FIPS-140-2 results of sequences $x$ and $y$ generated by the map~\ref{4}.}
	\begin{center}
		\begin{tabular}{c|p{105pt}|c c c}
			\hline
			Tests& 
			Sub-tests& 
			$ \mathbf{x} $& 
			$ \mathbf{y} $& 
			Decision \\
			\hline \hline
			\raisebox{-15.00ex}[0cm][0cm]{Runs test}& 
			P-value  & 
			0.3489& 
			0.1546& 
			Pass \\
			
			& 
			0 runs, length 1  & 
			2499& 
			2413& 
			Pass \\
			
			& 
			0 runs, length 2  & 
			1235& 
			1186& 
			Pass \\
			
			& 
			0 runs, length 3 & 
			612& 
			604& 
			Pass \\
			
			& 
			0 runs, length 4  & 
			305& 
			312& 
			Pass \\
			
			& 
			0 runs, length 5  & 
			152& 
			157& 
			Pass \\
			
			& 
			0 runs, length 6$+$& 
			153& 
			154& 
			Pass \\
			
			& 
			Longest run of 0& 
			14& 
			14& 
			Pass \\
			
			& 
			1 runs, length 1& 
			2592& 
			2489& 
			Pass \\
			
			& 
			1 runs, length 2  & 
			1287& 
			1257& 
			Pass \\
			
			& 
			1 runs, length 3  & 
			646& 
			645& 
			Pass \\
			
			& 
			1 runs, length 4  & 
			320& 
			334& 
			Pass \\
			
			& 
			1 runs, length 5  & 
			166& 
			170& 
			Pass \\
			
			& 
			1 runs, length 6$+$& 
			166& 
			172& 
			Pass \\
			
			& 
			Longest run of 1& 
			14& 
			13& 
			Pass \\
			\cline{1-5} 
			\raisebox{-2.00ex}[0cm][0cm]{Monobit test}& 
			P-value& 
			0.3104& 
			0.5489& 
			Pass \\
			
			& 
			No. of 1s 20000-bitstream& 
			10113& 
			10039& 
			Pass \\
			\cline{1-5} 
			\raisebox{-2.00ex}[0cm][0cm]{Poker test }& 
			p-value  & 
			0.1440& 
			0.1926& 
			Pass \\
			
			& 
			Y- value & 
			13.1008& 
			17.5232& 
			Pass \\
			\hline
		\end{tabular}
		\label{tab3}
	\end{center}
\end{table*}

\section{Chaos based image encryption and decryption}
\label{section:section4}
This section introduces a new  image encryption algorithm. Figure~\ref*{fig:diagram} illustrates the structure of the proposed algorithm, which achieves the confusion and diffusion processes by the hyperchaotic sequences and elliptic curve over the Galois field $GF_{2^8}$. Specifically, the enhanced Duffing map~\eqref{4} is used to generate hyperchaotic sequences for scrambling the pixels of a plain-image through image scrambling algorithm. Subsequently, the diffusion process is accomplished by field matrix, S-box, and hyperchaotic sequence. Simulations results demonstrate that the proposed encryption algorithm gives the users a flexibility to encrypt several kinds of images such as Grey scale, Medical, and RGB images with a higher level of security.

\subsection{Confusion based bit-plane transformation}
An efficient encryption algorithm should disassemble the high correlations between adjacent pixels. These high correlations can be de-correlated by scrambling adjacent pixels to different positions. To ensure an efficient scrambling process, we divided the plain-images into $ 8 $-bit-plane. Then the positions of all adjacent pixels are randomly scrambled by the image scrambling algorithm. The latter is demonstrated in  Algorithm \ref{Scram}. The algorithm typically illustrates the pseudo-code of scrambling and de-scrambling processes.

\begin{algorithm}
	\SetAlgoLined
	\caption{Image scrambling at bit level}
	\label{Scram}
	\KwIn{Plain-image of the size $ m\times n$.}
	\KwOut{The scrambled and de-scrambled images}
	Generate hyperchaotic sequences $\{x\}$ and $\{y\}$ with the long of $k$, where $k \geq m\times n $;\\
	Calculate $Sx=ceil(mod(x\times 10^5,256))$ and to form matrix  is of $Sx_{m\times n}\leftarrow Reshape(Sx,m,n)$\\
	Calculate $Sy=ceil(mod(y\times 10^14,m\times n))$ and to form a sequence where each element don't repeat and non zero\\
	$\%$ Image~scrambling\\
	Reshape $A_{m\times n}\longleftarrow \{a(i)\}$ for $i=1,2,...,m\times n$ and iterate $a(Sy(i))$\\
	Reshape $a(Sy(i))\longrightarrow A_{m\times n})$\\
	$\%$ Image~de-scrambling\\
	Calculate $inv.(Sy)$ from $Sy$\\
	Reshape $CI_{m\times n}\longleftarrow \{a(i)\}$ for $i=1,2,...,m\times n$ and iterate $a(inv(Sy(i)))$\\
	Reshape $a(inv(Sy(i)))\longrightarrow Decryptedimage_{m\times n}$\\
	Divide the plain-image into $8$-bit plane using  i.e. $A_{m\times n}$(plain-image)= $A_{m\times n\times 8}$   \\
	Divide $Sx$ into $Sx_{m\times n\times 8}$\\
	Initialize A matrix $B$ is of order $m\times n\times 8$ i.e. $B_{m\times n\times 8}=zeros(m,n,8)$\\
	\For{k=1 to $8$}{\For{j=1 to $n$}{\For{i=1 to $m$}{$B(i,j,k)= bitxor(A(i,j,k),Sx(i,j,k))$}}}
	\textbf{Return} ~$B$\\ 
	Set the bit position at each plain using the previous plane\\
	Initialize a scramble matrix $Sc^0=zeros(m,n)$\\
	\For{k=1 to $8$}{$Sc^k= bitset(B(:,:,k),Sc^{k-1})$}
	\textbf{Return} ~$Sc_{m\times n}$\\ 
\end{algorithm} 

\begin{algorithm}
	\SetAlgoLined
	\caption{The field matrix and its inverse.}
	\label{field}
	\KwIn{The initial conditions $IC_1$, and $IC_2$, as well as the primitive polynomial $f(x)$.}
	\KwOut{The field matrix and inverse field matrix. }
	Calculate $F(1)=mod(dec2bin(IC_1\times IC_2\times 10^3), f(x))$;\\
	\While{$i=1, i++$}
	{$F(i+1)=mod(dec2bin(F(i)\times IC_1\times IC_2\times 10^3),f(x))$
		{Calculate $F=[F(1), F(2),..., F(8)]$}\\
		\eIf {$det(F)\neq 0$ over $Z_2$}
		{ \Return $F$} {Calculate $F=[F(i+1),F(i+2),...,F(i+8)]$ }}
	Get $F_{8-bit\times 8-bit}$\\
	To calculate inverse field matrix\\
	Calculate $F(1)=mod(dec2bin(IC_1\times IC_2\times 10^3), f(x))$\\
	\While{$i=1, i++$}
	{$inv.F(i+1)=mod(dec2bin(inv.F(i)\times IC_1\times IC_2\times 10^3), f(x))$
		{Calculate $inv.F=[inv.F(1),..., inv.F(8)]$}\\
		\eIf {$det(F\times inv.F)= 1$ over $Z_2$}
		{ \Return $inv.F$} {Calculate $inv.F=[inv.F(i+1),..., inv.F(i+8)]$}}
	Get $inv.F_{8-bit\times 8-bit}$\\
\end{algorithm}


\begin{algorithm}
	\SetAlgoLined
	\caption{The S-Box and its inverse.}
	\label{S-box}
	\KwIn{The sequences $\{x_1\}$ and $\{y_1\}$, as well as the extract points $P$, field matrix $ F $ and $f(x)$.}
	\KwOut{The S-Box and inverse S-Box.}
	Calculate $Index^1_{256\times 256}\leftarrow reshape(x_1,256,256)$;\\
	Calculate $Index^2_{256\times 256}\leftarrow reshape(y_1,256,256)$;\\
	\For{j=1 to $256$}{\For{i=1 to $256$}{$Sb(i)\leftarrow Bin2Dec(mod(Dec2Bin(Index^1(P(i,1)\otimes Dec2Bin(P(1,j)),f(x))$;}}
	Calculate $SB_{16\times 16}\leftarrow reshape(SB,16,16)$;\\
	\For{j=1 to $16$}{\For{i=1 to $16$}{$S-Box(i,j)=Bin2Dec(mod([F]_{8\times 8}\times [Dec2Bin(SB(i,j)]_{8\times 1}),2)\oplus Dec2Bin(Index^2(16.i,16.j)))$;}}
	\textbf{Return} ~ S-Box;\\
	To calculate inverse S-Box \\
	Calculate $inv[F]\leftarrow inverse ~of ~ [F]_{8\times 8} ~ over~ Z_2$;\\
	\For{j=1 to $16$}{\For{i=1 to $16$}{$inv.(S-Box)(i,j)=Bin2Dec(mod(inv.[F]_{8\times 8}\times [Dec2Bin(Index^2(16.i,16.j)]_{8\times 1}),2)\oplus Dec2Bin(SB(i,j)))$;}}
	\textbf{Return} ~ Inv.(S-Box);
\end{algorithm}  
\subsection{S-box and the field matrix}
We construct the  S-box and the field matrix based on the points of an elliptic curve over the Galois field $GF_{2^n}$. The description of the elliptic curve and its points is given below. 
The elliptic curve is a set of points that satisfy the following Weierstrass equation:
\begin{equation}
E: y^2 + a_1xy +a_3 y = x^3 +a_2x^2 +a_4x +a_6,
\end{equation} 
where  $a_1, a_2, a_3, a_4, a_5$, and $a_6$ represent the parameters and the initial conditions of the map~\eqref{4}  $A,\alpha, B, \beta, IC_1$, and $IC_2$, respectively.

The elliptic curve equation over the field $GF_{2^8}$ is defined as
\begin{equation}
E_{2^8}(a,b): y^2+xy=x^3+ax+b,  \label{eq-elliptic}
\end{equation}
which is obtained by the following transformation
\begin{equation*}
\left( x, y\right) \rightarrow \left( a_1^2x+\frac{a_3}{a_1}, a_1^3y+\frac{a_1^2a_4+a_3^2}{a_1^3} \right), 
\end{equation*}
{
	where $a_1\neq 0$ and $a,b \in F_{2^8}$. Thus, the curve is defined over the field $GF_{2^8}$, and $a$ and $b$ are calculated as
	\begin{equation}
	\begin{split}
	a=floor(mod((a_3/(a_1^3)+a_2/a_1^2)\times 10^5,256)),\\
	b=floor(mod((c_2^2+c_1^3+c_1^2+c_1+a_6+a_4.c_1)\times 10^5,256)),
	\end{split}
	\end{equation}
	with $c_1=a_3/a_1^3$ and $c_2=(a_3^2.a_4+a_3^3)/a_1^6$.}
\par 
To construct the field matrix $ F $, we  first extract some points  which satisfy Eq. \eqref{eq-elliptic}, by means of the following primitive polynomial.
\begin{equation*}
f(x)=1+x^2+x^3+x^4+x^8.
\end{equation*}
Here, if the generator $g$ satisfies $f(g)=0$, one obtains  $g^8=g^4+g^3+g^2+1$, where $ g \equiv(00000010)$.
Thus, the properties of the points on elliptic curve $E_{2^8}(a,b)$ over the field $F_{2^8}$ can be stated as follows:
\begin{itemize}
	\item For any point  $P(x_p,y_p)\in E_{2^8}(a, b)$ satisfying Eq. \eqref{eq-elliptic} and if $O$  is the additive identity, i.e., $P+O=P$  then $-P=(x_p,x_p+y_p)$ and $P-P=O$. 
	\item If $P=(x_p,y_p)$, $Q=(x_q,y_q)\in E_{2^8}(a, b)$ and $P\neq \pm Q$, then $R=P+Q=(x_r,y_r)$, where $(x_r,y_r)=(\alpha^2+\alpha+x_p+x_q+A,\alpha(x_p+x_r)+x_r+y_p)$ and $\alpha={(y_q+y_p)}/{(x_p+x_q)}$. Again, if $P=Q$, then $2P=(x_r,y_r)=(\lambda^2+\lambda+A,x_p^2+(\lambda+1)x_r$, where $\lambda=x_p+\frac{y_p}{x_p}$.
\end{itemize}
The algorithm for the field matrix and its inverse is stated in Algorithm \ref{field}. The field matrix is constructed by the initial conditions $IC_1$, and $IC_2$ of the enhanced map \eqref{4} and the primitive polynomial $f(x)$. It is imperative to note here that the field matrix is invertible over $Z_2$, which gives the possibility to generate the inverse S-Box for decryption process. 
\par
\begin{figure}[h!]
	
	\includegraphics[width=8cm, height=4cm]{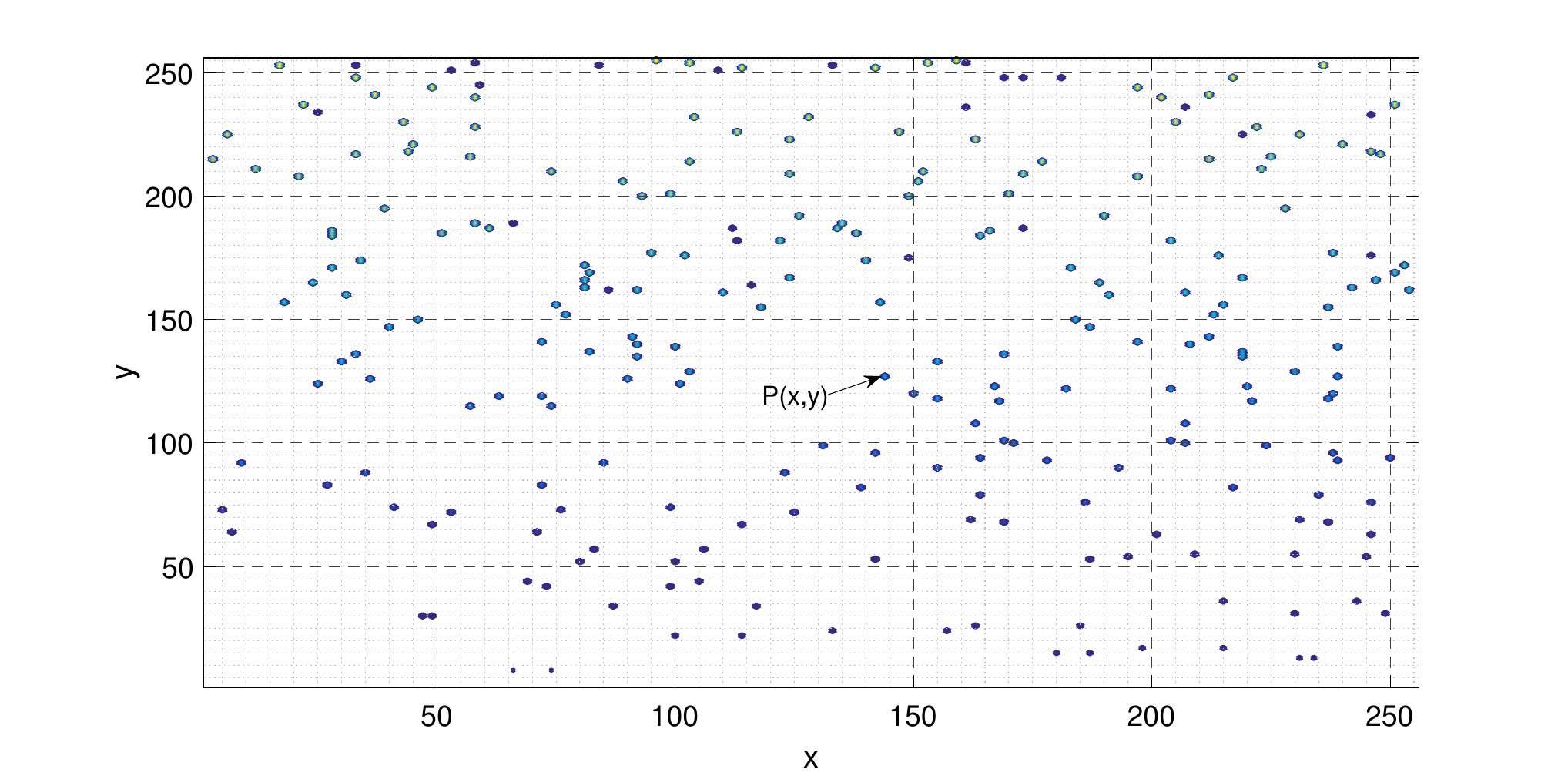}
	\caption{{Extracted points on the elliptic curve $E_{2^8}(a,b)$ over the field $GF_{2^8}$} are shown}
	\label{fig:points}
	
\end{figure}

\begin{figure*}
	\centering
	\includegraphics[width=16cm, height=8cm]{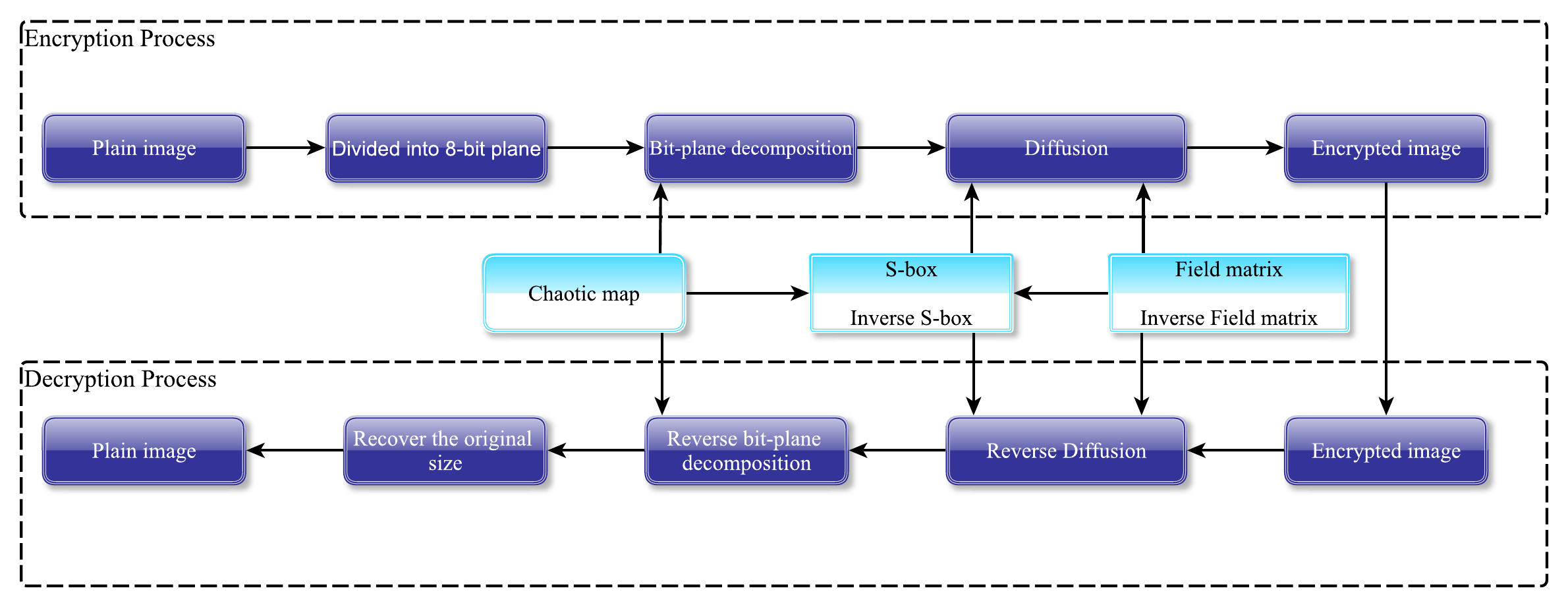}
	\centering
	\caption{Schematic diagram of the proposed image encryption and decryption algorithm.}
	\label{fig:diagram}
\end{figure*}

 
Next, in order to extract points $P(x,y)$ to form the S-box of order $16\times 16$, we require $256$ number of points, i.e.   $P=[P_1,P_2,...,P_{256}]$  that satisfy Eq. \eqref{eq-elliptic}. As an illustration,  we consider the parameter  values as  $A=15.02154872$, $B=3.50142547$, $\beta=5.02314723$, $\alpha =10.00148752$, $IC_1=0.2501254781$, $IC_2=0.2712548731$ so that $a=g^7+g^5+g^3+1$ and $b=g^7+g^6+g^3+g^2$. The extracted  points $P(x,y)$ on the elliptic curve $E_{2^8}(a,b)$ are  shown in Fig. \ref{fig:points}.
In what follows, we consider the  control parameters and the initial conditions that give rise the hyperchaotic states of the  enhanced Duffing map \eqref{4}. The  hyperchaotic sequences $\{x\}$ and $\{y\}$  so generated are then transformed over the Galois field $GF_{2^8}$ into two new sequences as follows:
\begin{flalign*}
\begin{cases}
\begin{aligned}
x_1=&floor(mod(x\times 10^5,256)),\\
y_1=&floor(mod(y\times 10^5,256)).
\end{aligned}
\end{cases}
\end{flalign*}
The new sequences $\{x\}$ and $\{y\}$ along with the extracted points $ P(x, y)$  of the  elliptic curve   over the Galois field $GF_{2^8}$, the primitive   polynomial $ f(x) $ and the field matrix $ F $ are then used to construct the S-box. The corresponding algorithm is  given in Algorithm~\ref{S-box}.  

\begin{figure*}
	\begin{center}
		\begin{subfigure}[h]{0.235\textwidth}
			\includegraphics[width=4.5cm, height=3.6cm]{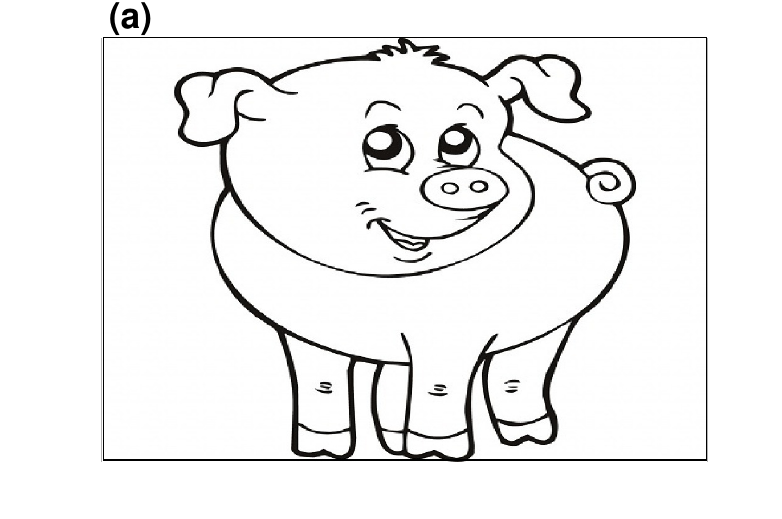}
		\end{subfigure}
		\begin{subfigure}[h]{0.235\textwidth}
			\includegraphics[width=4.5cm, height=3.6cm]{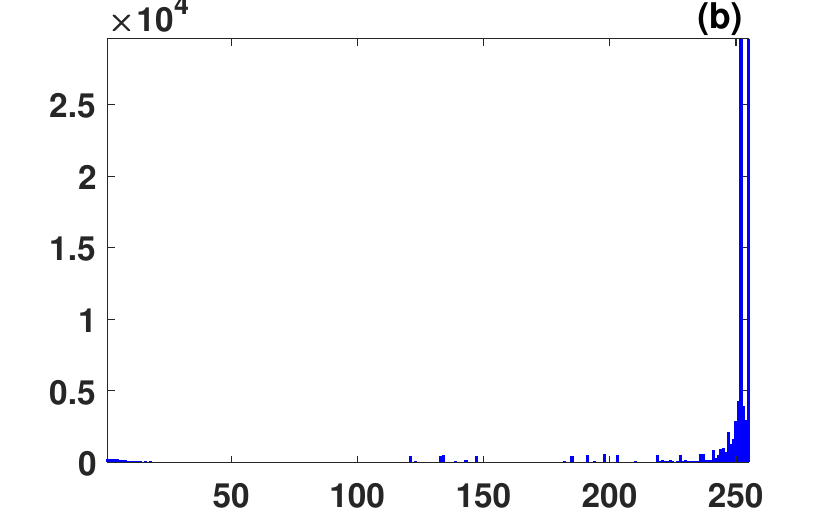}
		\end{subfigure}
		\begin{subfigure}[h]{0.235\textwidth}
			\includegraphics[width=4.5cm, height=3.6cm]{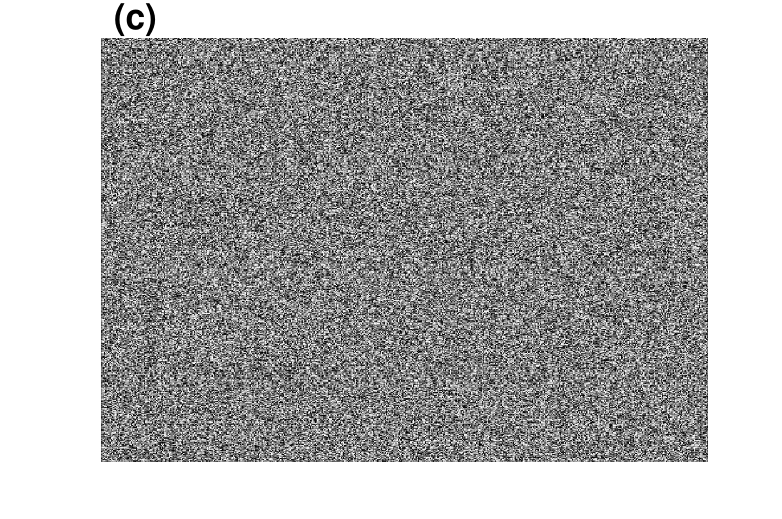}
		\end{subfigure}
		\begin{subfigure}[h]{0.235\textwidth}
			\includegraphics[width=4.5cm, height=3.6cm]{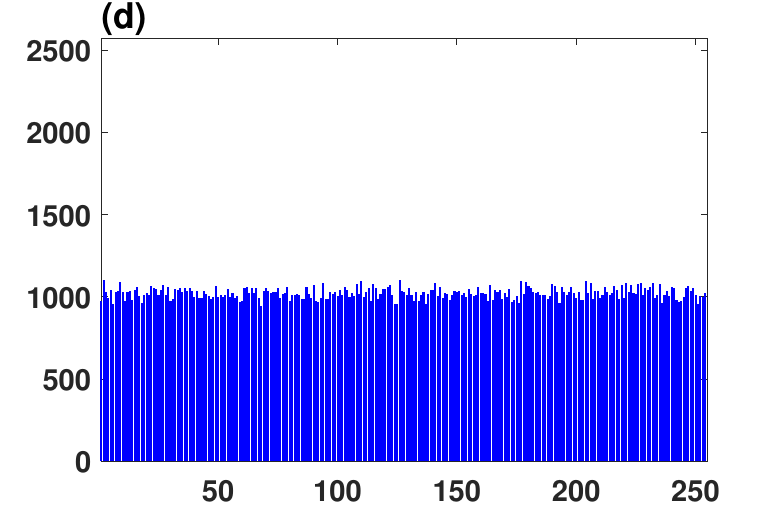}
		\end{subfigure}
		
		\begin{subfigure}[h]{0.235\textwidth}
			\includegraphics[width=4.5cm, height=3.6cm]{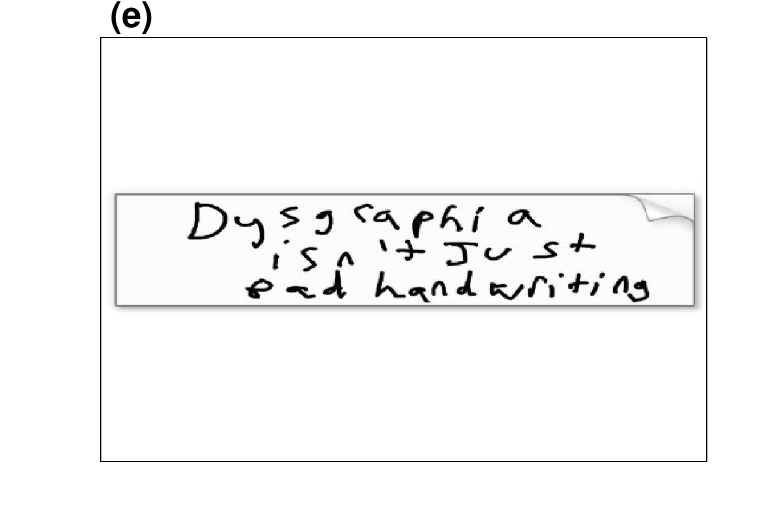}
		\end{subfigure}
		\begin{subfigure}[h]{0.235\textwidth}
			\includegraphics[width=4.5cm, height=3.6cm]{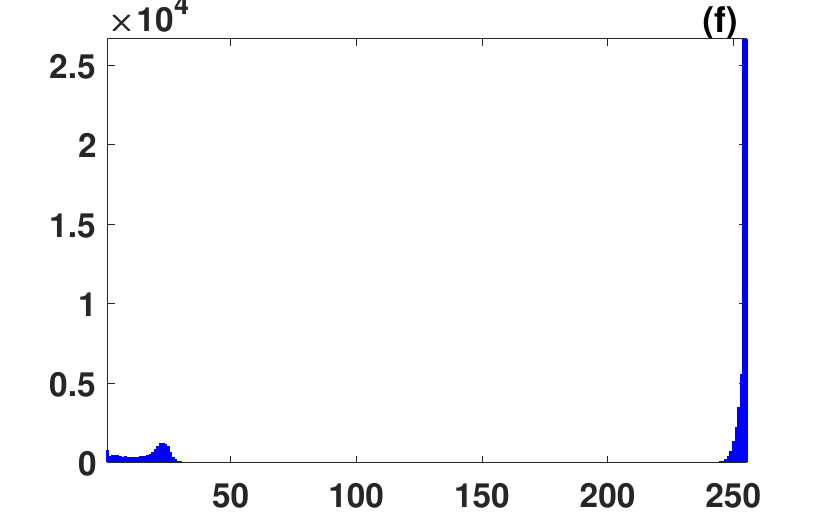}
		\end{subfigure}
		\begin{subfigure}[h]{0.235\textwidth}
			\includegraphics[width=4.5cm, height=3.6cm]{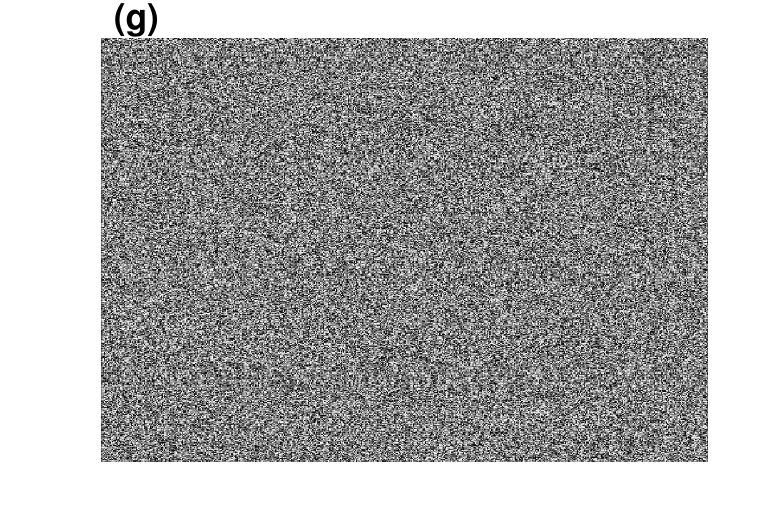}
		\end{subfigure}
		\begin{subfigure}[h]{0.235\textwidth}
			\includegraphics[width=4.5cm, height=3.6cm]{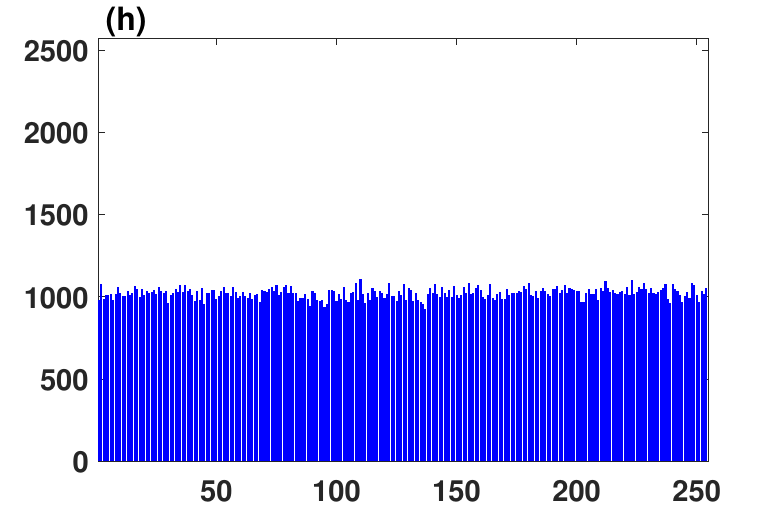}
		\end{subfigure}
		
		\begin{subfigure}[h]{0.235\textwidth}
			\includegraphics[width=4.5cm, height=3.6cm]{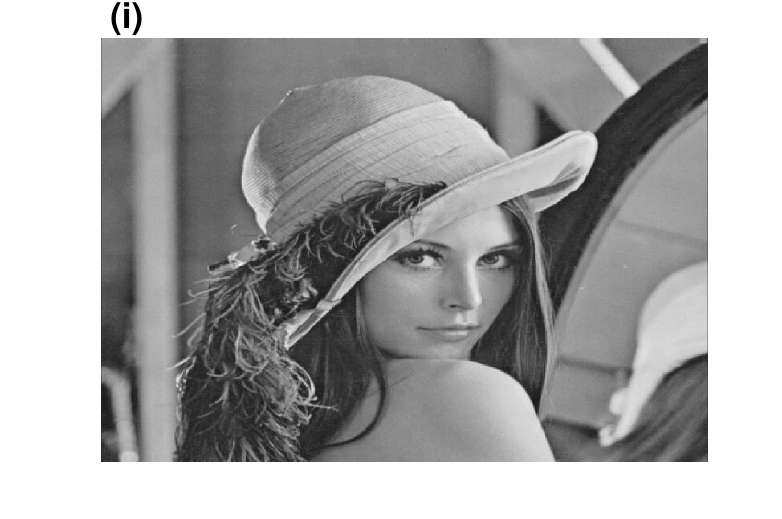}
		\end{subfigure}
		\begin{subfigure}[h]{0.235\textwidth}
			\includegraphics[width=4.5cm, height=3.6cm]{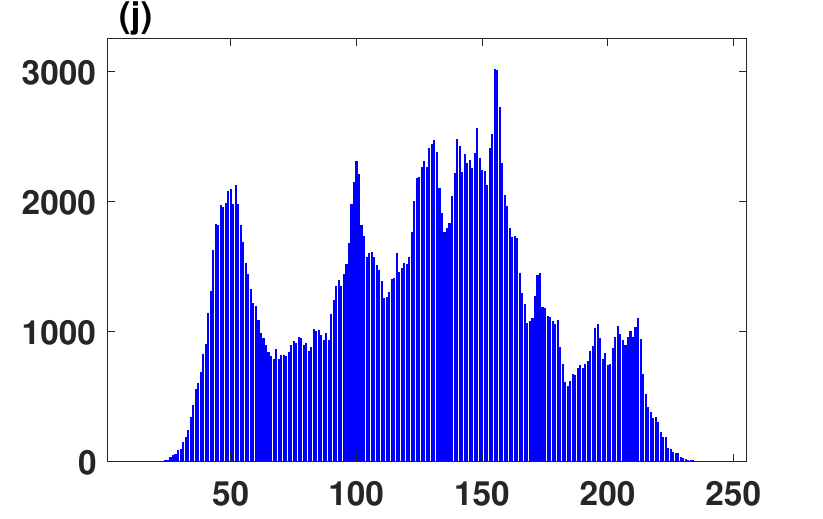}
		\end{subfigure}
		\begin{subfigure}[h]{0.235\textwidth}
			\includegraphics[width=4.5cm, height=3.6cm]{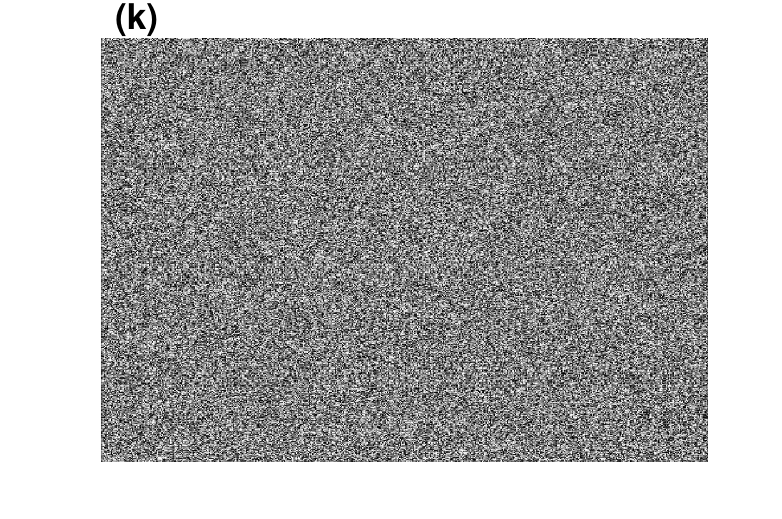}
		\end{subfigure}
		\begin{subfigure}[h]{0.235\textwidth}
			\includegraphics[width=4.5cm, height=3.6cm]{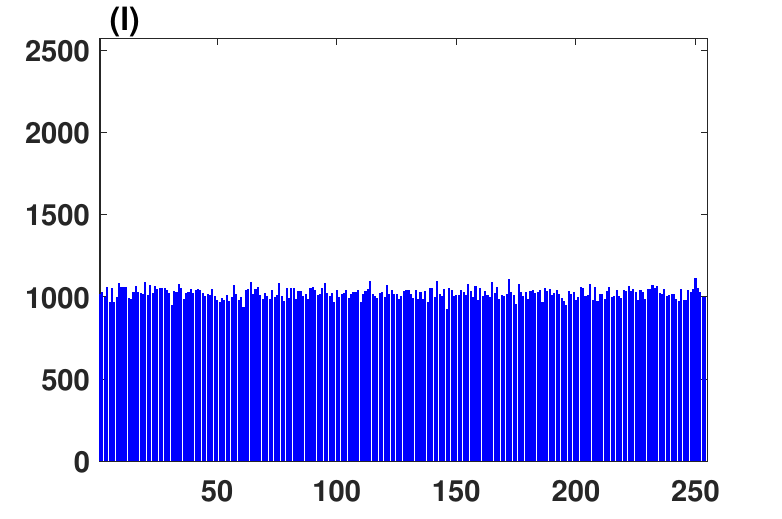}
		\end{subfigure}
		
		\begin{subfigure}[h]{0.235\textwidth}
			\includegraphics[width=4.5cm, height=3.6cm]{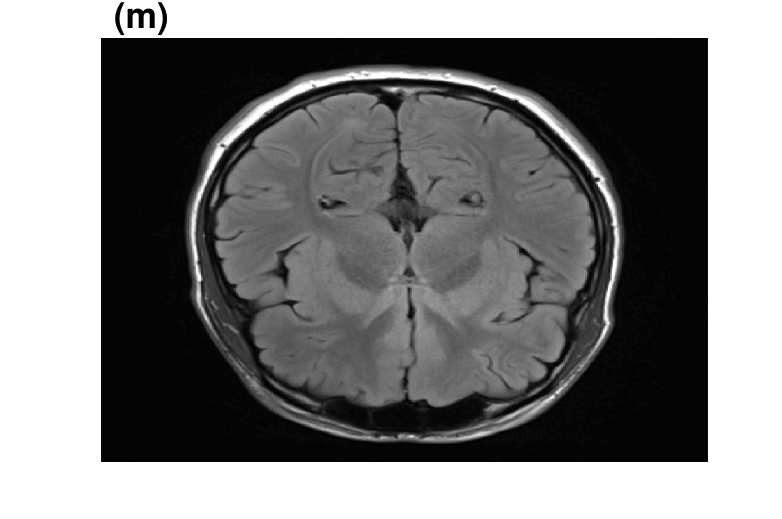}
		\end{subfigure}
		\begin{subfigure}[h]{0.235\textwidth}
			\includegraphics[width=4.5cm, height=3.6cm]{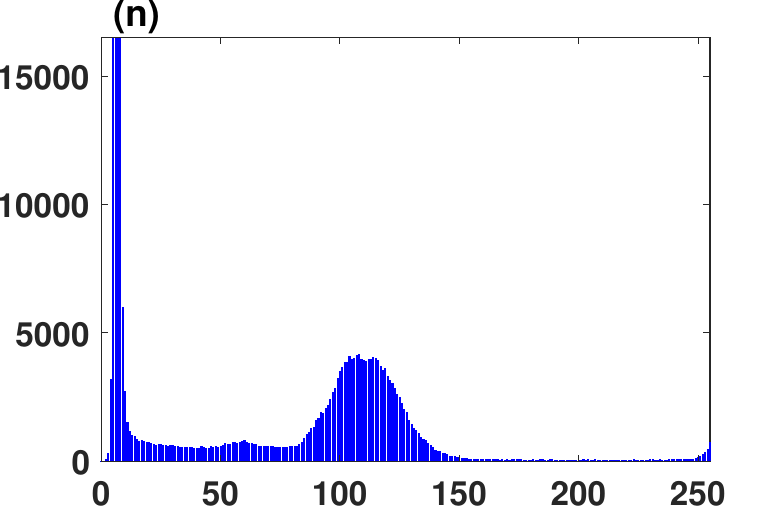}
		\end{subfigure}
		\begin{subfigure}[h]{0.235\textwidth}
			\includegraphics[width=4.5cm, height=3.6cm]{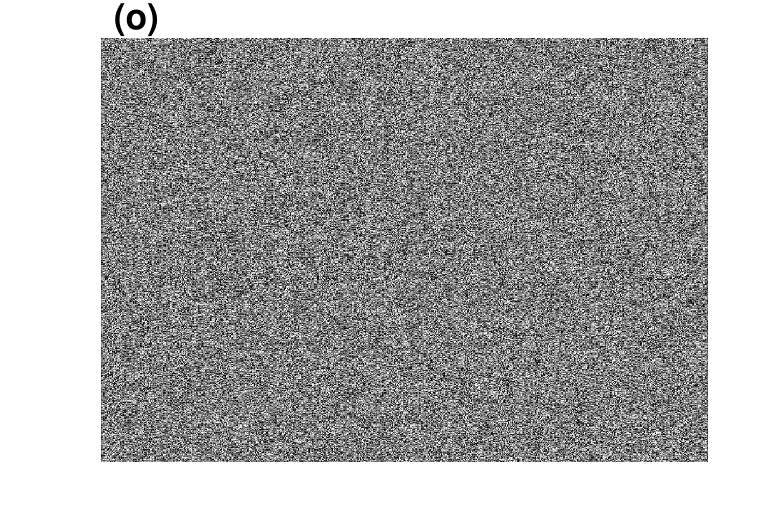}
		\end{subfigure}
		\begin{subfigure}[h]{0.235\textwidth}
			\includegraphics[width=4.5cm, height=3.6cm]{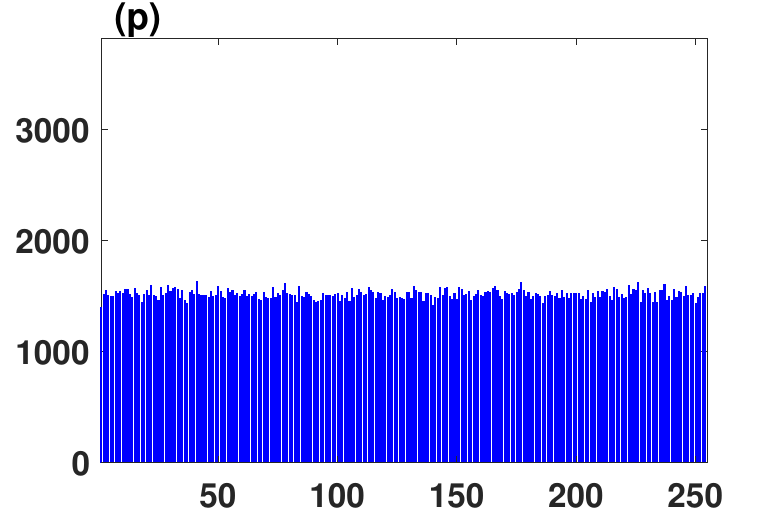}
		\end{subfigure}
		
		\begin{subfigure}[h]{0.235\textwidth}
			\includegraphics[width=4.5cm, height=3.6cm]{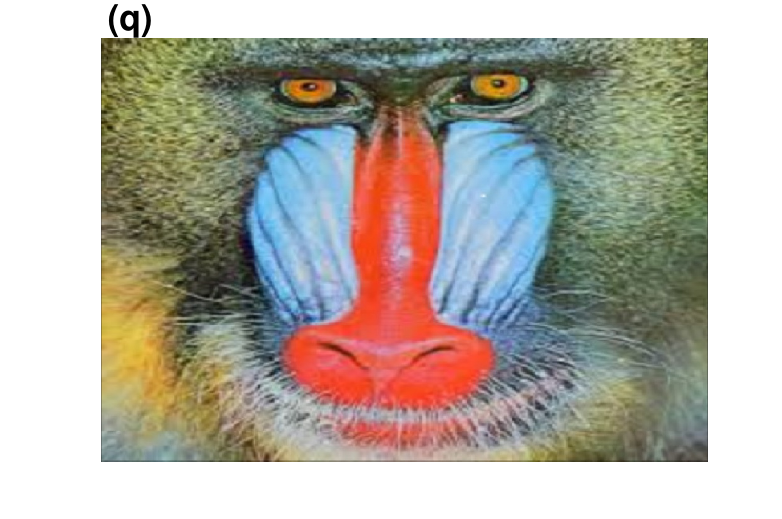}
		\end{subfigure}
		\begin{subfigure}[h]{0.235\textwidth}
			\includegraphics[width=4.5cm, height=3.6cm]{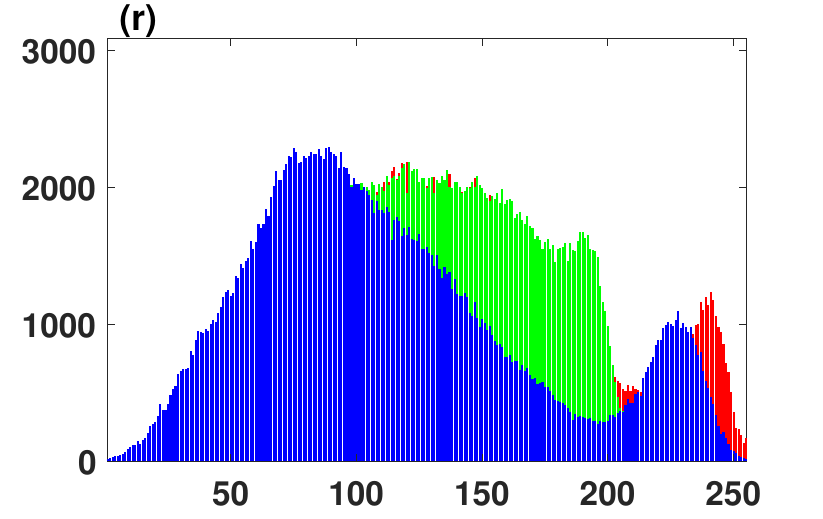}
		\end{subfigure}
		\begin{subfigure}[h]{0.235\textwidth}
			\includegraphics[width=4.5cm, height=3.6cm]{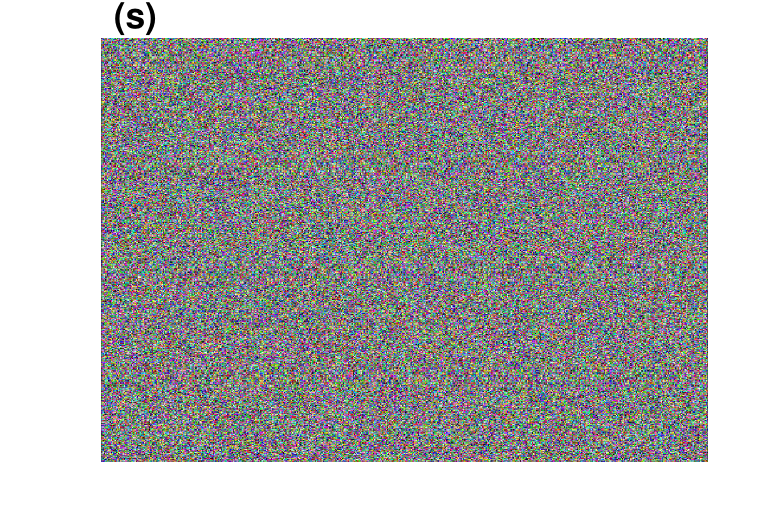}
		\end{subfigure}
		\begin{subfigure}[h]{0.235\textwidth}
			\includegraphics[width=4.5cm, height=3.6cm]{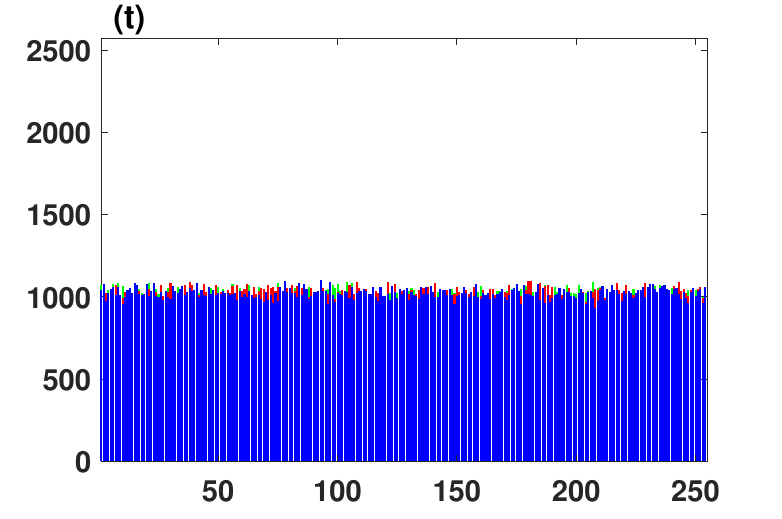}
		\end{subfigure}
		
		\begin{subfigure}[h]{0.235\textwidth}
			\includegraphics[width=4.5cm, height=3.6cm]{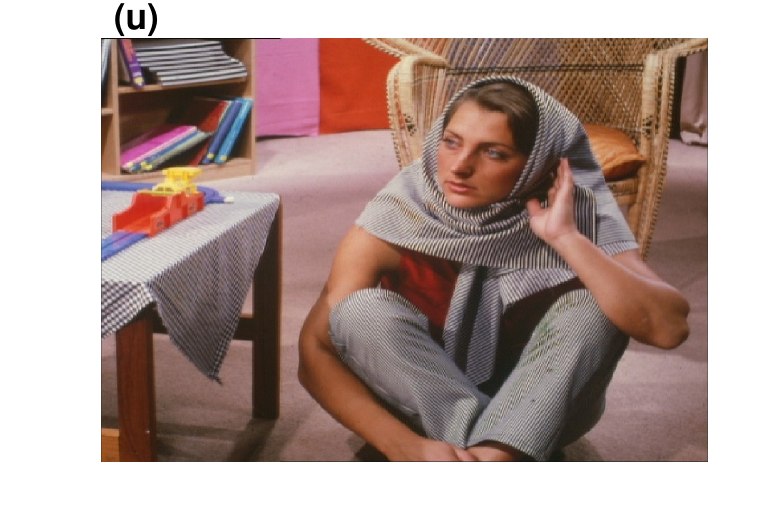}
		\end{subfigure}
		\begin{subfigure}[h]{0.235\textwidth}
			\includegraphics[width=4.5cm, height=3.6cm]{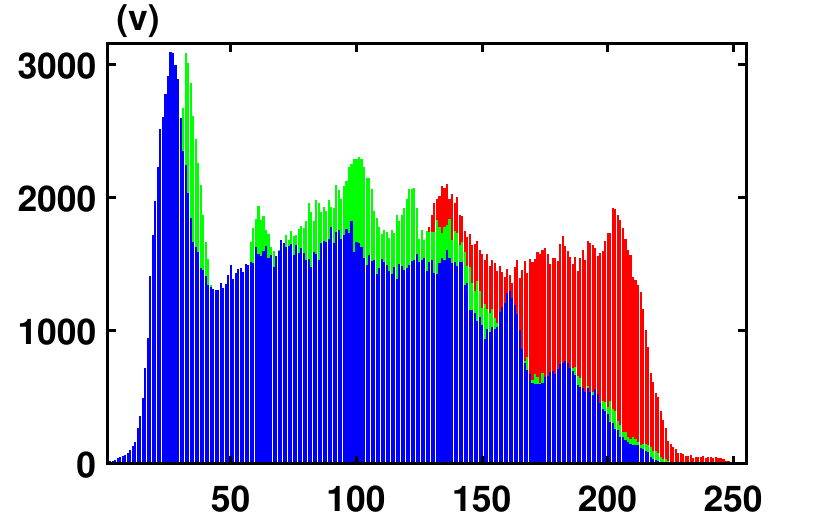}
		\end{subfigure}
		\begin{subfigure}[h]{0.235\textwidth}
			\includegraphics[width=4.5cm, height=3.6cm]{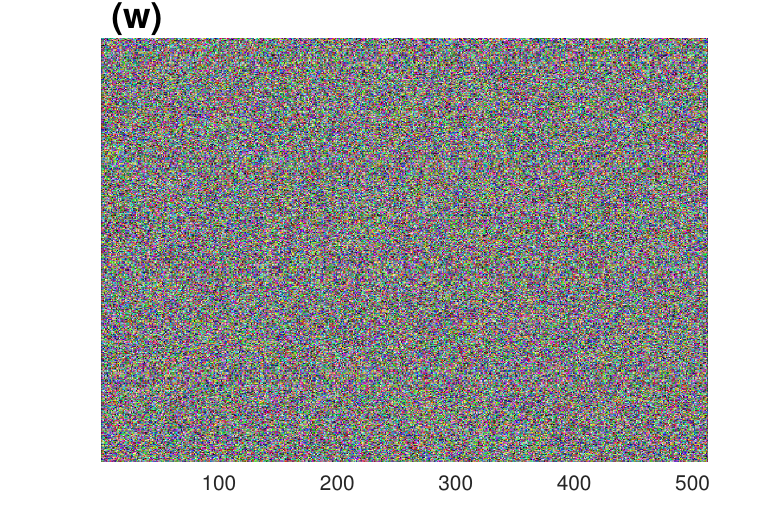}
		\end{subfigure}
		\begin{subfigure}[h]{0.235\textwidth}
			\includegraphics[width=4.5cm, height=3.6cm]{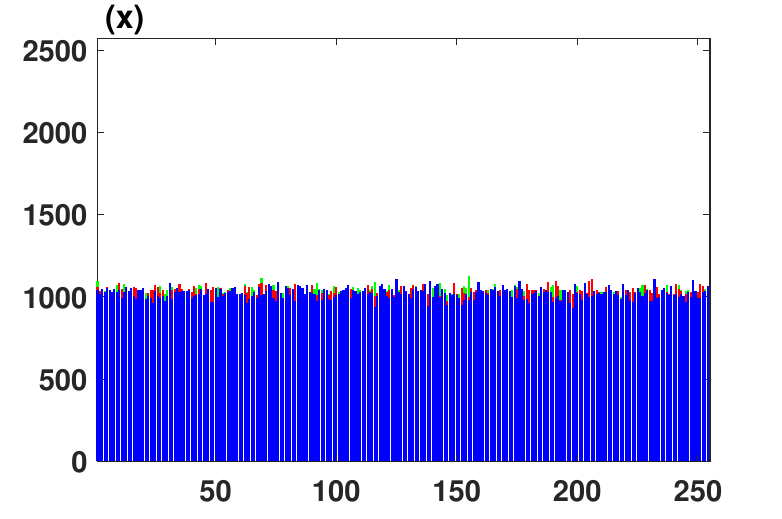}
		\end{subfigure}
	\end{center}
	\caption{Encryption results of different images: the first column depicts the plain-images including animal sketch image, hand writing image, Lena grey-scale image, medical image, animal color image, Barbara color image; the second column depicts histograms of the plain-images; the third column depicts the encryption of the plain-images; the fourth column depicts histograms of the encrypted images.}
	\label{fig:fig6}
\end{figure*}

\subsection{Diffusion process}
An image encryption algorithm has the ability to defeat chosen-plaintext attack when it has an efficient diffusion process. Therefore, this section introduces a new  algorithm based on the field matrix $ F $, S-Box, and the hyperchaotic sequence $\{y\} $. The   process is described as follows:
\begin{itemize}
	\item[Step 1 :] The pixel values of $Sc_{m\times n}$ are divided  into $Sc^i_{k\times k}$, where $k=1, 2,..., 16$, and $i=1, 2,..., \frac{m\times n}{K^2}$.
	
	\item[Step 2 :] Using the generated hyperchaotic sequence $\{y\}$ by the enhanced map, $Sy\longleftarrow$ $ceil(mod(y\times 10^5, 256))$ is   calculated.
	
	\item[Step 3 :] The new scramble image matrix $SC^i_{k\times k}$ is generated by the field matrix, which can be defined as $SC^i_{k\times k} \longleftarrow Bin2Dec(mod([F]_{8\times 8}\times Dec2Bin(Sc^i_{k\times k}(i_1, j_1, l)), 2)$, where  $i_1, j_1=1, 2,...k$ and $l=1,  2,..., i$.

	\item[Step 4 :] The block cipher image (CI) $CI^i_{k\times k}$ is obtained by the bitwise XOR operation among $SC^i_{k\times k}$, S-Box and the sequence $S_{y}$.

	\item[Step 5 :] The cipher image is obtained by reshaping $CI^i_{k\times k}$ into $CI_{m\times n}$ as:  $CI_{m\times n} \longleftarrow reshape(CI^i_{k\times k}, m, n)$.
\end{itemize} 

\subsection{The process of decryption}
The receiver section gets the cipher image along with the initial condition and control parameters of the enhanced map~\eqref{4}. Then the receiver constructs the field matrix $F$ and the S-box, which are considered as secrete key. Using the elliptic curve, hyperchaotic sequences, and the secrete key, the receiver can construct the inverse field matrix and the inverse S-Box  which represent the original key.   Finally, the plain-image is recovered by the  following decryption process. 

\begin{itemize}
	\item[Step 1 :] The  cipher image   $CI_{m\times n}$ is divided  into $CI^i_{k\times k}$ for $k=1,2,...,16$ and $i=1,2,...,\frac{m\times n}{K^2}$.
	\item[Step 2 :] Having obtained the inverse field matrix ($inv.(F)$),  the chaos sequence $\{y\}$ from the Duffing map   and  $Sy\longleftarrow ceil(mod(y\times 10^5,256))$, calculate $inv.(Sy)\longleftarrow inv.(F)\times Sy^i_{k\times k}$ for $k=1,2,...,16$ and $i=1,2,...,\frac{m\times n}{k^2}$, and construct the inverse S-Box using the algorithm \ref{S-box}.
	\item[Step 3 :] Obtain the new cipher block matrix   as $CI'^i_{k\times k} \longleftarrow Bin2Dec(mod(inv.[F]_{8\times 8}\times Dec2Bin(CI^i_{k\times k}(i_1,j_1,l)),2)$ for $i_1,j_1=1,2,...k$ and $l=1,2,...,i$ with $k=1,2,..16$ and $i=1,2,...,\frac{m\times n}{k^2}$. 
	\item[Step 4 :] Calculate the scramble image matrix  using the bitwise XOR operation as\\ $SC^i_{k\times k}(k1,k2,i2)\longleftarrow mod((mod((inv.(S-Box(k1,k2))\oplus CI'^i_{k\times k}(k1,k2,i2)),256)\oplus\\ inv.Sy^i_{k\times k}(k1,k2,i2)),256)$ for $k_1,k_1=1,2,...k$ and $i_2=1,2,...,i$ with $k=1,2,..16$  and  $i=1,2,...,\frac{m\times n}{k^2}$. 
	\item[Step 5 :] Reshape the scramble image as  \\ $SC_{m\times n} \longleftarrow reshape(SC^i_{k\times k},m,n)$.
	\item[Step 6 :] Descramble the  scramble  image   using the same Algorithm \ref{Scram} to recover the plain image.
\end{itemize}
\section{Simulation results and key analysis}
\label{section:section5}
In this section, the proposed image encryption algorithm is simulated to demonstrate its efficiency. Moreover, the robustness of the employed key is investigated by performing the key space and the key sensitivity analyses.
\subsection{Encrypting different kinds of images}
To illustrate the ability of the proposed image encryption algorithm for ciphering different types of images, Figure~\ref{fig:fig6} depicts the encryption results with uniformly distributed histograms of various kinds of plain-images including Grey scale, RGB, Sketch, and Hand writing images with the size of ($512\times 512$). Moreover, this figure shows the encryption of a medical image   of  size ($630\times 630$) with its histogram. It can be seen from these results that the proposed encryption algorithm can effectively encrypt various kinds of images.

\subsection{Key space analysis}
Typically, the security key of chaos-based cryptography contains two main components, namely the initial conditions and the control parameters of the employed chaotic map. In the proposed encryption algorithm, the parameters and the initial conditions of the enhanced Duffing map~\eqref{4} are  the main roots of the secrete keys. Each parameter and initial values are considered  with $ 15 $ to $ 16 $ decimal places, which means that the complexity of each parameter and  the initial value is $ 2^{52} $. Besides that, the hyperchaotic sequences of the enhanced map~\eqref{4} are generated by the parameters and the initial conditions for constructing the field matrix and the S-box. So, the key space in producing   $F$ and the S-Box is $2^{64} (\rm{for}~ 64-\rm{bit}~ \rm{operation})\times 2^{102}(\rm{initial} ~\rm{conditions}) = 2^ {166}$, and the total key combinations is $2^{478}$ i.e., the size of the key in the proposed algorithm is $478$ bits. Consequently, the security key of the proposed algorithm achieves the standard requirement \cite{alvarez2006some}. 

Furthermore, the total time to break an encrypted image is calculated as follows 
\begin{equation*}
Y= \frac{T\times 1000}{FLOPS}\times 3153600, 
\end{equation*}
where $ Y $ is the total years to break an encrypted image, and $ T $ is the total security key space. A super computer  has $10^{15}$  floating-point operation per second (FLOPS). So, the total time to break the encrypted image by the proposed algorithm is approximately $1.230593278\times 10^{137}$ years.

\section{Security analysis}
\label{section:section6}
The most important indicator to evaluate an image encryption algorithm is the security performance of its encrypted image. This section introduces an analysis framework to investigate the security of the encrypted images by the proposed algorithm.

\subsection{Robustness analysis of noise and data loss}
\begin{figure*}
	\centering
	\includegraphics[width=15cm, height=12cm]{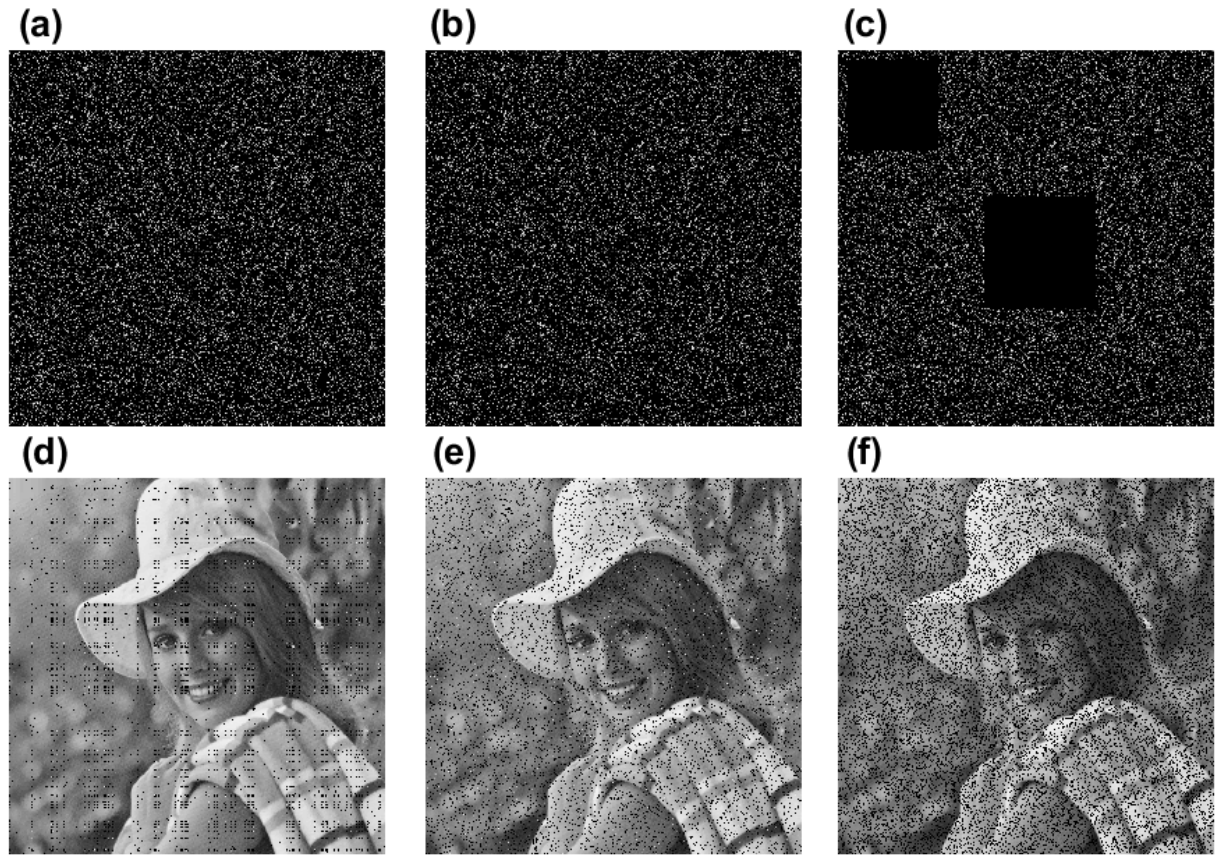}
	\centering
	\caption{Robustness analysis of noise and data loss: (a) and (d) depict the encrypted image with $5\%$ Gaussian noise, and the decrypted image  respectively; (b) and (e) depict the encrypted image with $12\%$ salt and pepper noise, and the corresponding decrypted image  respectively; (c) and (f) depict the encrypted image with $15\%$ data loss, and the decrypted image  respectively.}
	\label{fig:fig9}
\end{figure*}

Different kinds of noise and data lose can corrupt the encrypted images. So, the image encryption algorithms should be able to resist these kinds of noise and data lose. The first and second columns of Figure~\ref{fig:fig9} demonstrate the quality results of the recovered image when the corresponding encrypted image undergoes Gaussian noise with $5\%$ density, as well as salt and pepper noise with $12\%$ density. It can be observed that although the encrypted images have noise, the corresponding recovered images contain the most visual information of the original images. Besides that, the proposed algorithm has successfully recovered the images with  Peak Signal-to-Noise Ratio (PSNR) equal to $22.32$ and Mean Squre Error (MSE) equal to $381.08$ even with an addition of  $5\%$ Gaussian noise. Meanwhile, the obtained PSNR and MSE for addition of $12\%$ of salt and pepper noises are  $17.3648$ and $1192$  respectively. Moreover, the third column of Figure~\ref{fig:fig9} shows that the recovered images by the proposed algorithm can  still be recognizable when the encrypted image has $ 15\% $ data loss.  As a result, the proposed encryption algorithm can resist different kids of noise and data loss.

\begin{table*}[h]
	\caption{The NPCR and UACI values are calculated with $\alpha = 0.05$ for different images using some existing schemes and our scheme.}
	\setlength{\extrarowheight}{3pt}
	\begin{tabular}{p{40pt}p{4pt}|p{25pt}|p{45pt}p{40pt}p{40pt}p{34pt}p{34pt}p{33pt}}
		\hline
		\multicolumn{1}{c|}{Index} & 
		\multicolumn{2}{|c|}{Image size} & 
		Our scheme& 
		LICM\cite{cao2018novel}& 
		ICMIE\cite{cao2020designing}& 
		Zhou\cite{zhou2013image}& 
		Wu\cite{wu2012image}& 
		Liao\cite{liao2010novel}\\
		\hline
		\hline
		\raisebox{-3.00ex}[0cm][0cm]{NPCR}& 
		\multicolumn{2}{|c|}{$256\times256\ge99.5693\% $} & 
		\multicolumn{1}{c}{6/6}& 
		\multicolumn{1}{c}{6/6}& 
		\multicolumn{1}{c}{6/6}& 
		\multicolumn{1}{c}{6/6}& 
		\multicolumn{1}{c}{6/6}& 
		\multicolumn{1}{c}{0/6} \\
		
		&\multicolumn{2}{|c|}{$512\times 512  \ge 99.5893\% $ } & 
		\multicolumn{1}{c}{18/18}& 
		\multicolumn{1}{c}{17/18}& 
		\multicolumn{1}{c}{18/18}& 
		\multicolumn{1}{c}{17/18}& 
		\multicolumn{1}{c}{17/18}& 
		\multicolumn{1}{c}{1/18} \\
		
		&\multicolumn{2}{|c|}{$1024\times 1024  \ge 99.5994\% $} & 
		\multicolumn{1}{c}{3/3}& 
		\multicolumn{1}{c}{3/3}& 
		\multicolumn{1}{c}{3/3}& 
		\multicolumn{1}{c}{3/3}& 
		\multicolumn{1}{c}{2/3}& 
		\multicolumn{1}{c}{0/3} \\
		\hline 
		Pass Rate & 
		\multicolumn{2}{c}{} & 
		\multicolumn{1}{c}{27/27} & 
		\multicolumn{1}{c}{26/27} & 
		\multicolumn{1}{c}{27/27} & 
		\multicolumn{1}{c}{26/27} & 
		\multicolumn{1}{c}{25/27} & 
		\multicolumn{1}{c}{1/27} \\
		\hline
		\hline
		\raisebox{-3.00ex}[0cm][0cm]{UACI}& 
		\multicolumn{2}{|c|}{$256\times 256\  (33.2824-33.6447)$ } & 
		\multicolumn{1}{c}{6/6}& 
		\multicolumn{1}{c}{6/6}& 
		\multicolumn{1}{c}{6/6}& 
		\multicolumn{1}{c}{1/6}& 
		\multicolumn{1}{c}{6/6}& 
		\multicolumn{1}{c}{0/6} \\
		
		& 
		\multicolumn{2}{|c|}{$512\times 512\  (33.3730-33.5541)$ } & 
		\multicolumn{1}{c}{18/18}& 
		\multicolumn{1}{c}{16/18}& 
		\multicolumn{1}{c}{18/18}& 
		\multicolumn{1}{c}{4/18}& 
		\multicolumn{1}{c}{15/18}& 
		\multicolumn{1}{c}{0/3} \\
		
		& 
		\multicolumn{2}{|c|}{$1024\times 1024\  (33.4183-33.5088)$} & 
		\multicolumn{1}{c}{3/3}& 
		\multicolumn{1}{c}{2/3}& 
		\multicolumn{1}{c}{3/3}& 
		\multicolumn{1}{c}{2/3}& 
		\multicolumn{1}{c}{1/3}& 
		\multicolumn{1}{c}{0/3} \\
		\hline 
		Pass Rate &\multicolumn{2}{c}{} & 
		\multicolumn{1}{c}{27/27}& 
		\multicolumn{1}{c}{24/27}& 
		\multicolumn{1}{c}{27/27}& 
		\multicolumn{1}{c}{7/27}& 
		\multicolumn{1}{c}{22/27}& 
		\multicolumn{1}{c}{0/27} \\
		\hline
	\end{tabular}
	\label{tab4}
\end{table*}

\subsection{Differential attack resistance}
A vulnerable encryption scheme could attack by observing the change in the encrypted images when a small change or modification happens in the corresponding plain image. This type of attack is namely chosen plaintext attack or differential attack. The NPCR and UACI tests could be employed to estimate the resistance of encryption algorithms against differential attacks. The NPCR indicates the pixel change rate, while the UACI indicates the unified averaged changed intensity. These two measures are as follows: 
\begin{equation}
NPCR =\dfrac{ \sum_{i, j}^{m, n}D(i, j)}{m\times n} \times 100, 
\end{equation}  
where $D(i,j)$ is the change of the pixel values from the plain-image to the encrypted image due to the encryption process, in which
\begin{equation}
D(i,j)= \left\lbrace \begin{array}{ccc}  
0 & when &  P(i, j)=CI(i, j) \\
1  & when &  P(i, j)\neq CI(i, j). 
\end{array} \right.
\end{equation}
As the pixel values are changed, the UACI can be used to determine the average intensity of the difference between the original and the encrypted image, where
\begin{equation}
UACI= \dfrac{1}{m\times n} \sum_{i, j}^{m, n} \dfrac{|P(i, j)-CI(i, j)|}{255} \times 100. 
\end{equation}
However, Wu \cite{wu2011npcr} presented a new standard of NPCR and UACI measures for better estimating the ability of an encryption algorithm for resisting differential attack. In this regard, an image encryption algorithm can pass the NPCR test when its NPCR value is bigger than a level $\alpha$, which is given by the following equations.
\begin{flalign}
\begin{split}
N_{\alpha}^{*}=&\mu_{N}- \Phi^{-1}(\alpha)\sqrt{\frac{F}{255}},\\
N_{\alpha}^{*}=&\frac{F-\Phi^{-1}(\alpha)\sqrt{\frac{F}{255}}}{F+1},
\end{split}
\end{flalign}
where $\Phi^{-1}$ is the inverse CDF of the standard Normal distribution $ N(0, 1) $, and $ F $ is the largest supported pixel value compatible with the cipher text image format. An encryption algorithm   successfully passes the UACI test when the simulation value is in the range of $UACI\in (U_{\alpha}^{-}, U_{\alpha}^{+})$. Here, $ U_{\alpha}^{-} $ and $ U_{\alpha}^{+} $ are given by
\begin{flalign}
U_{\alpha}^{-}=&\mu_U-\Phi^{-1}(\alpha/2)\sigma_U, \nonumber\\
U_{\alpha}^{+}=&\mu_U+\Phi^{-1}(\alpha/2)\sigma_U,
\end{flalign}
with 
\begin{flalign}  
\sigma_U=&\sqrt{\frac{(F+2)(F^2+2F+3)}{18(F+1)^2 \times 255F}}, \nonumber \\
\mu_U=&\frac{F+2}{3F+3}.
\end{flalign}
\begin{figure*}
	\begin{center}
		\begin{subfigure}[h]{0.24\textwidth}
			\includegraphics[width=4.5cm, height=5cm]{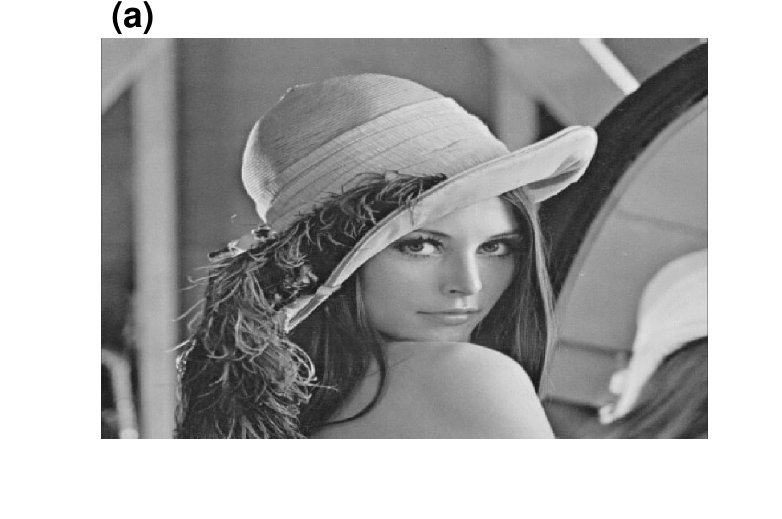}
		\end{subfigure}
		\begin{subfigure}[h]{0.24\textwidth}
			\includegraphics[width=4.5cm, height=5cm]{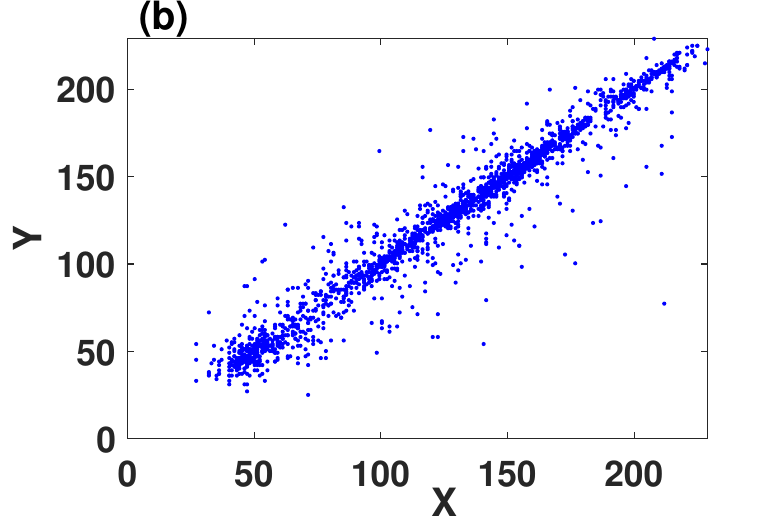}
		\end{subfigure}
		\begin{subfigure}[h]{0.24\textwidth}
			\includegraphics[width=4.5cm, height=5cm]{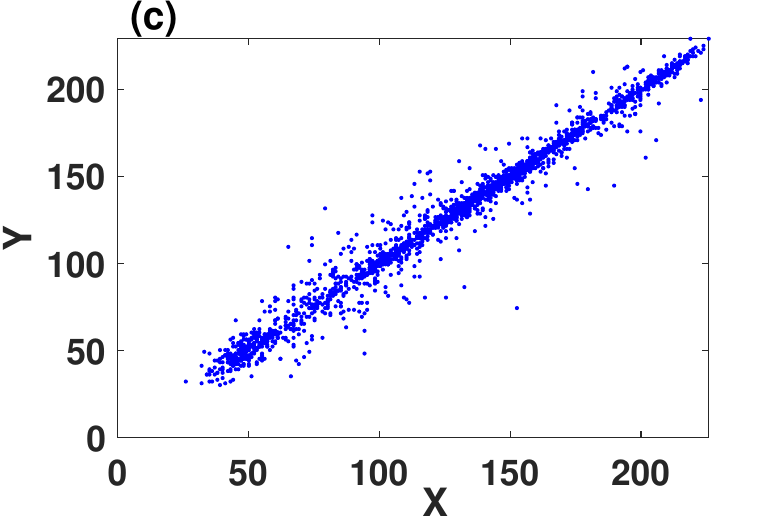}
		\end{subfigure}
		\begin{subfigure}[h]{0.24\textwidth}
			\includegraphics[width=4.5cm, height=5cm]{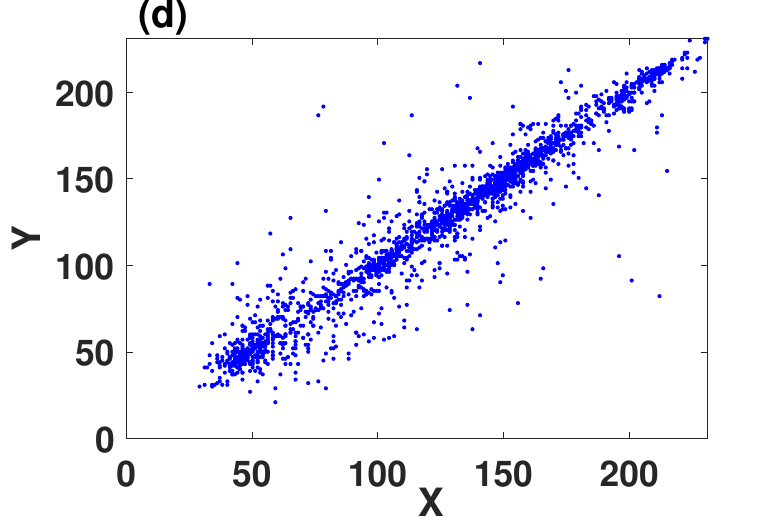}
		\end{subfigure}
		
		\begin{subfigure}[h]{0.24\textwidth}
			\includegraphics[width=4.5cm, height=4.8cm]{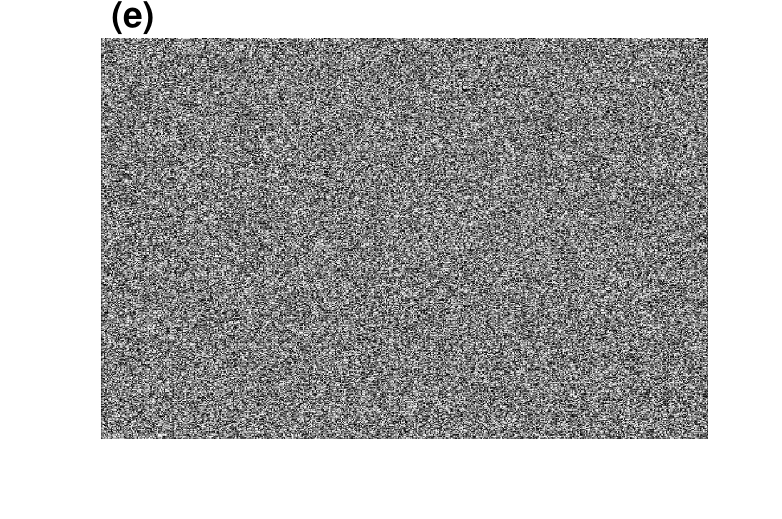}
		\end{subfigure}
		\begin{subfigure}[h]{0.24\textwidth}
			\includegraphics[width=4.5cm, height=4.8cm]{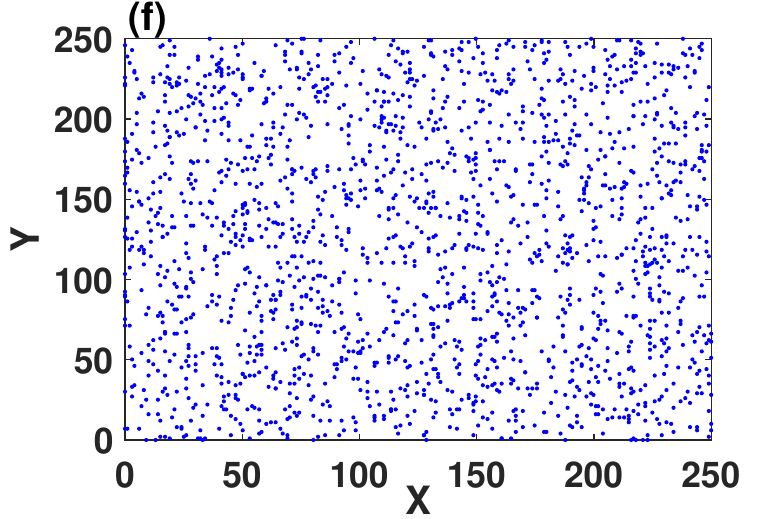}
		\end{subfigure}
		\begin{subfigure}[h]{0.24\textwidth}
			\includegraphics[width=4.5cm, height=4.8cm]{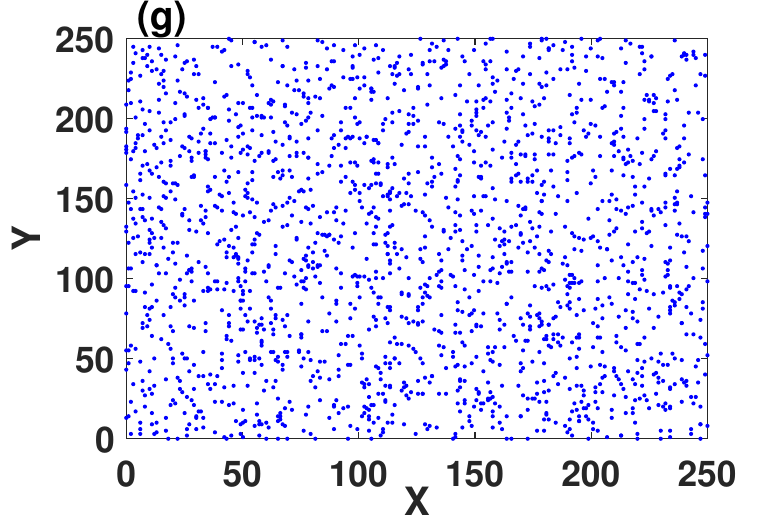}
		\end{subfigure}
		\begin{subfigure}[h]{0.24\textwidth}
			\includegraphics[width=4.5cm, height=4.8cm]{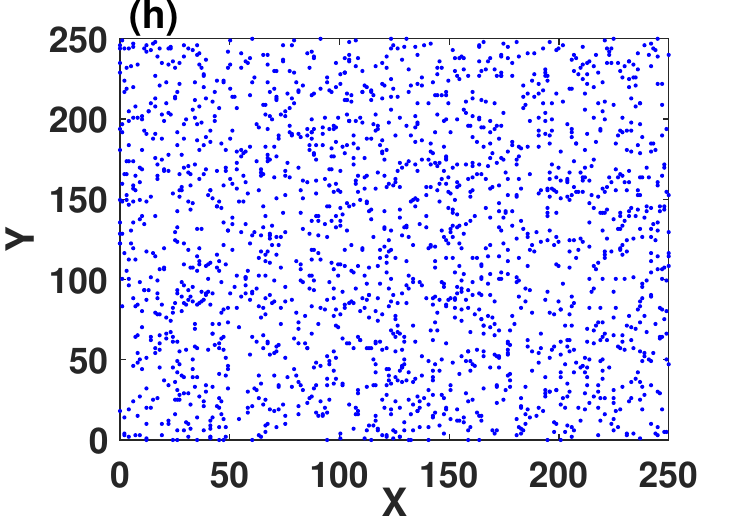}
		\end{subfigure}
	\end{center}
	\caption{The correlation of the neighboring pixel pairs: the first column depicts the plain-image and the corresponding encrypted image; the second column depicts the horizontal direction of plain-image and the encrypted image; the third column depicts the vertical direction of the plain-image and the encrypted image; the fourth column depicts the diagonal direction of the plain-image and the encrypted image.}
	\label{fig:fig8}
\end{figure*}
In this test, for each plaint image, namely, $ PI_{1} $, we generate an another image, namely, $ PI_{2} $ by selecting a pixel from $ PI_{1} $ and changing its value by 1-bit. Subsequently, the UPCR and UACI values can be calculated by generating the encrypted images of both $ PI_{1} $ and $ PI_{2} $. The NPCR and UACI results of $27$ different types of images, which are taken from USC-SIPI Miscellaneous data set and which have been encrypted by the proposed algorithm and several other schemes, are illustrated in Table~\ref{tab4}.
From these results it is found that that the proposed algorithm has superior or competitive performance in defending the differential attack. 
\subsection{Resisting Noise Attack Analysis}
	The encoded image version is inevitably exposed to different
	types of noises, when the data passes through a real communication
	channel. This noise can cause problems during the
	acquisition of the original image. Therefore, the algorithm
	should be noise resistant, so that the encryption scheme can
	be valid. The Peak Signal-to-Noise Ratio (PSNR) is used to
	measure the quality of the decoded image after the attacks. For the image components, PSNR can be obtained by the
	following formulation
	\begin{equation}
	PSNR=10\times log_{10}(\frac{255\times 255}{MSE})~DB
	\end{equation}
	where
	\begin{equation}
	MSE= \frac{1}{m\times n} \sum_{i=1}^{m}\sum_{j=1}^n\parallel (i,j) - DCI(i,j)\parallel
	\end{equation}
	MSE is the mean square error between the original and recovered images and is represented as P and DCI respectively, with the size of $m\times n$. The first and second columns of Fig. \ref{fig:fig9} demonstrate the quality results of the recovered image when the corresponding encrypted image undergoes Gaussian noise with $5\%$ density, as well as salt and pepper noise with $12\%$ density.The MSE and PSNR of these decoded images are shown in Table \ref{tablemse}. From this table and Fig. \ref{fig:fig9}, we can understand that the original image is entirely obtained again, which is noticeable, the PSNR value is near about $30 dB$ when Gaussian noise density $1\%$, $22 dB$ with $5\%$ density and $18 dB$ for $12\%$ density and the decoded images are highly correlated with original image.
	
\begin{table*}
\caption{MSE and PSNR analysis of plain image and decrypted image in which cipher image is introduced with Gassian Noise with different level of density $1\%$, $5\%$ and $12\%$.}
\begin{center}
			\begin{tabular}{|p{0.6in}|c|c|p{0.5in}|c|c|c|c|} \hline 
			\multirow{2}{*}{Image}  & \multirow{2}{*}{Noise Density}  & \multicolumn{3}{|c|}{MSE} & \multicolumn{3}{|c|}{PSNR(dB)} \\ \cline{3-8} 
			&  & R & G & B & R & G & B \\ \hline 
			\multirow{3}{*}{Barbara}  & 0.01 & 105.2502 & 111.2145 & 108.23 & 28.8560 & 28.1520 & 27.9289 \\  
			& 0.05 & 381.0801 & 378.5205 & 382.8906 & 22.3578 & 22.1245 & 22.1345 \\  
			& 0.12 & 1192.2452 & 1188.3542 & 1194.2578 & 17.3648 & 17.2548 & 17.2045 \\ \hline 
			\multirow{3}{*}{Lena} & 0.01 & 102.5625 & 103.2541 & 104.3252 & 29.7824 & 30.2569 & 29.1245 \\  
			& 0.05 & 370.2356 & 369.8567 & 372.5687 & 23.5689 & 23.5478 & 23.7482 \\  
			& 0.12 & 1178.2586 & 1174.6583 & 1180.8563 & 18.3562 & 18.6524 & 18.5242 \\ \hline 
			\multirow{3}{*}{Animal} & 0.01 & 104.1625 & 105.5141 & 106.2152 & 29.1824 & 29.1269 & 29.1451 \\  
			& 0.05 & 375.1256 & 374.1057 & 377.1067 & 22.1068 & 22.1054 & 22.4082 \\  
			& 0.12 & 1198.2045 & 1201.3212 & 1199.1057 & 16.4648 & 16.2554 & 16.3445\\ \hline
		\end{tabular}\label{tablemse}
\end{center}
	\end{table*}
	
	\begin{table*}
		\caption{The encrypted and decrypted NPCR and UACI values for different image  sizes are shown due to a small change in the parameters and/or initial condition.}
		\setlength{\extrarowheight}{3pt}
		\begin{tabular}{ c c c c} 
			\hline
			Index  & Image size  &Encrypted using actual initial conditions&	Decrypted using wrong  initial conditions \\
			\hline \hline
			\multirow{3}{*}{NPCR} & $256\times256 \ge99.5693\%$ & $99.655151367\%$&$99.588012695\%$ \\
			& $512\times 512 \ge 99.5893\%  $ &$99.633407592\%$&$99.623489379\%$\\ 
			&$1024\times 1024  \ge 99.5994\% $& $99.612331390\%$&$99.608325958\%$\\ 	
			\multirow{3}{*}{UACI} & $256\times 256 \  (33.2824-33.6447)$ &$33.353662490\%$&$33.47766058\%$\\
			& $512\times 512\  (33.3730-33.5541$ & $33.492037854\%$&$33.485890501\%$\\
			& $1024\times 1024\  (33.4183-33.5088)$ &$33.500159085\%$&$33.501732840\%$\\
			\hline
		\end{tabular}
		\label{tabkeysen}
	\end{table*}

\subsection{Key sensitivity analysis}
\begin{figure*}
	\begin{center}
		\begin{subfigure}[h]{0.32\textwidth}
			\includegraphics[width=6cm, height=5cm]{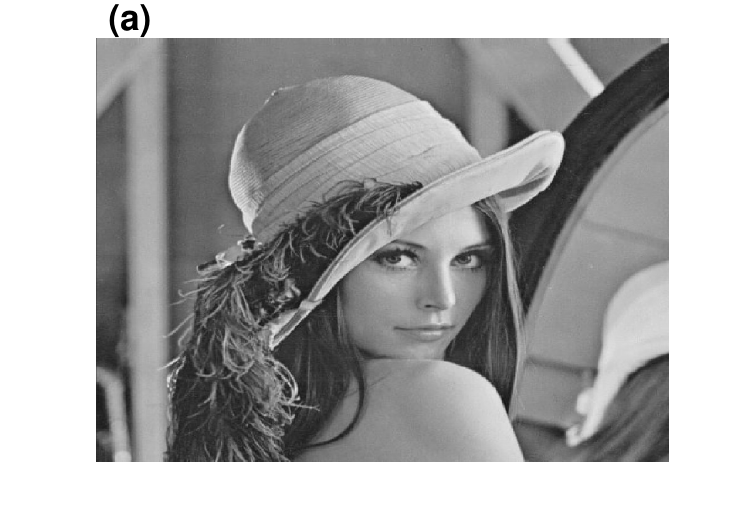}
		\end{subfigure}
		\begin{subfigure}[h]{0.32\textwidth}
			\includegraphics[width=6cm, height=5cm]{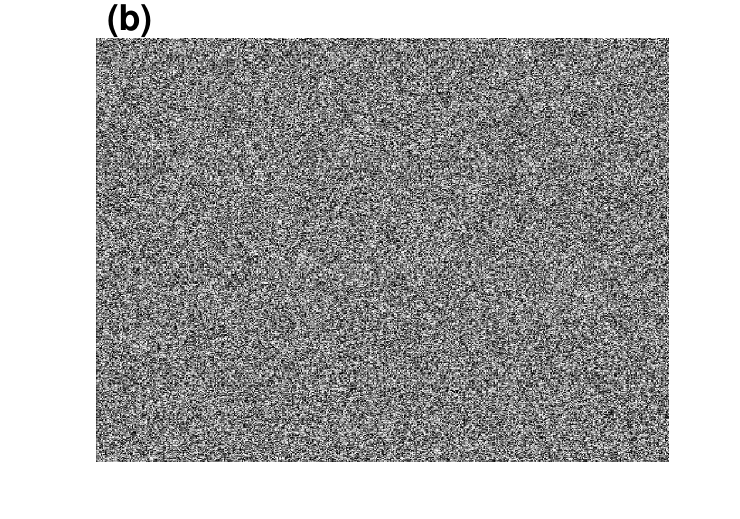}
		\end{subfigure}
		\begin{subfigure}[h]{0.32\textwidth}
			\includegraphics[width=6cm, height=5cm]{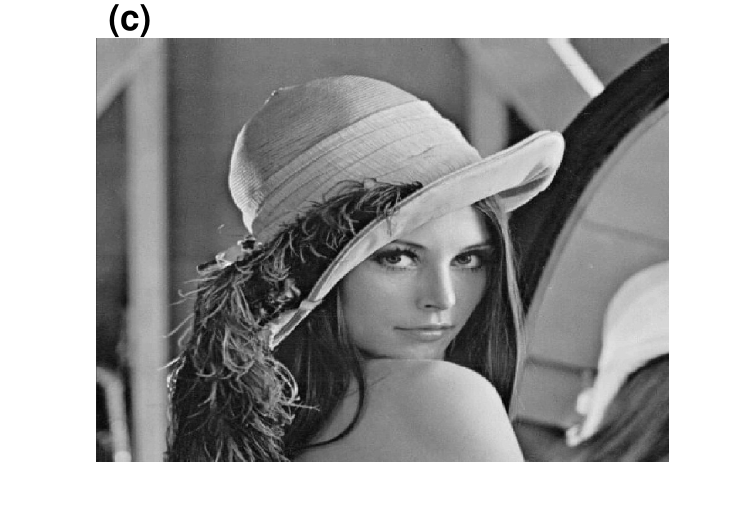}
		\end{subfigure}
		
		\begin{subfigure}[h]{0.32\textwidth}
			\includegraphics[width=6cm, height=5cm]{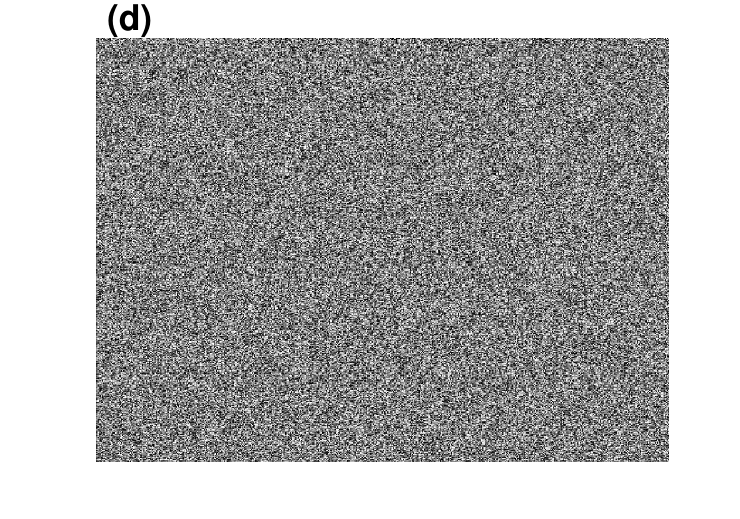}
		\end{subfigure}
		\begin{subfigure}[h]{0.32\textwidth}
			\includegraphics[width=6cm, height=5cm]{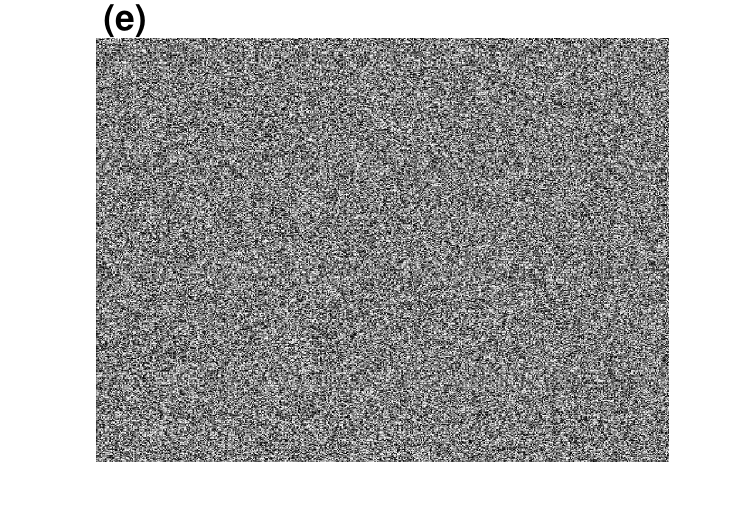}
		\end{subfigure}
		\begin{subfigure}[h]{0.32\textwidth}
			\includegraphics[width=6cm, height=5cm]{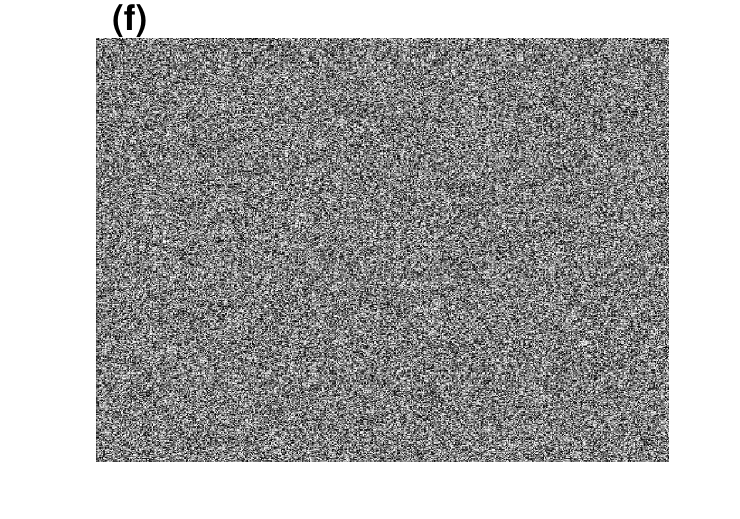}
		\end{subfigure}
	\end{center}
	\caption{Key sensitivity analysis: (a) plain-image, (b) the encrypted image, (c) the decrypted image with the right key.  Subplots (d)-(f) are  the decrypted images with  wrong keys. They are obtained   due to  a slight change in    both the parameters and the initial conditions [subplot (d)], in the parameters only [subplot (e)],  in the initial condition only [subplot (f)].}
	\label{fig:fig10}
\end{figure*}

The employed key of an image encryption scheme is considered as highly sensitive when the encrypted image cannot be recovered due to a slight difference in one of the key components. To visualize the key sensitivity of the proposed encryption algorithm, we set the main root of the secrete  keys, which represent the initial conditions and parameters of the map~\eqref{4}, as $A=15.02154872$, $B=3.50142547$, $\beta=5.02314723$, $\alpha =10.00148752$, $IC_1=0.2501254781$, $IC_2=0.2712548731$. Subsequently, we change  $14^{th}$ decimal places in the parameters, or initial conditions, or both to obtain three other keys. Figure~\ref{fig:fig10} demonstrates the key sensitivity in the decryption process with the original key and the modified keys.{ We consider different image sizes and calculate the NPCR and UACI values corresponding to the  encryption with    the actual parameters and  initial condition as well as the decryption with wrong parameters and initial condition. The results are   given in Table \ref{tabkeysen}.   Thus, from   Table \ref{tabkeysen} it is clear that even with a small change, i.e., a change in the $14$-th decimal place  of the parameters or initial condition,    the change in the pixel values is more than $99\%$  after decryption with wrong initial condition and/or parameters.}

\subsection{Correlation analysis}
The adjacent pixels of the original image are highly correlated along the vertical, diagonal, and horizontal directions. Thus, an efficient encryption algorithm can resist a statistical attack when the adjacent pixels of its encrypted image is nearly zero. To calculate the pixels correlation, let us first define the covariance between a pair of  pixel values $x$ and $y$, which is given by
\begin{equation}
Cov(x, y) = E[(x-E(x))(y-E(y))],
\end{equation}
where $E(x)$ and $E(y)$ are the means. Now, the correlation coefficients can be calculated by
\begin{equation}
\rho_{xy}=\dfrac{Cov(x, y)}{\sigma(x)\sigma(y)},~ ~\sigma(x),~\sigma(y)\neq0,
\end{equation}
where $\sigma(x)$ and $\sigma(y)$ are the standard  deviations of the distribution of the pixel.

In our analysis, adjacent pixels in horizontal, vertical, and diagonal directions are randomly chosen from both plaint and encrypted images, as shown in \ref{fig:fig8}. As can be seen, most of the pixels are close to the diagonal line of axis for the plain image. Meanwhile, the pixels of the encrypted image distribute randomly on the whole space. Furthermore, quantitative and comparison results of adjacent pixels correlations of Lena image, which is encrypted by the proposed encryption algorithm and other existing schemes, are illustrated in Table~\ref{tab2}. Clearly, the $\rho_{xy}$ values of our scheme are more closer to $ 0 $, and superior and competitive than those of some other schemes.

\begin{table*}
	\caption{The correlation coefficients of the plain-image and the encrypted image by various methods.}
	\setlength{\extrarowheight}{3pt}
\begin{center}
		\begin{tabular}{ c c c c c c c c} 
		\hline
		Index & Lena image & Our scheme & LICM\cite{cao2018novel}  & ICMIE\cite{cao2020designing} & 2D-LCCM\cite{Nan2022} & Xu\cite{xu2016novel} & Liao\cite{liao2010novel} \\ 
		\hline \hline
		Horizontal & 0.971921627 & -0.0009 & 0.0019 & -0.0008 & -0.0009 & 0.0230 & 0.0127\\ 
		
		Vertical & 0.9865777  &  0.0015 & 0.0012 & -0.0013 &-0.0005 & 0.0019 & -0.0190\\ 
		
		Diagonal & 0.96064343 &  -0.0010 & 0.0009 & 0.0018 & 0.0029 & 0.0034 & -0.0012\\
		\hline
	\end{tabular}
\end{center}
	\label{tab2}
\end{table*}

\begin{table*}
	\caption{The local Shannon entropy analysis with $\alpha=0.001$ $k=30$, $T_B=1936$ of images which are collected from USC-SIPI Miscellaneous data-set}.
\begin{center}
		\begin{tabular}{c c c c c c c}
		\hline 
		{Image name}& 
		Our scheme & 
		Wu \cite{wu2012image}& 
		Wang\cite{zhou2013image} & 
		Liao \cite{liao2010novel}& 
		Shen\cite{Shen2022}& 
		Kumar\cite{Kumar2022}\\
		\hline \hline
		5.1.09& 
		7.902212& 
		7.901985& 
		7.899212 & 
		7.904191&  
		7.997181& 
		7.999602 \\
		
		5.1.10& 
		7.901902& 
		7.902731 & 
		7.901125& 
		7.902371&  
		7.997282& 
		7.999626 \\
		
		5.1.11& 
		7.902425& 
		7.902446 & 
		7.901521&  
		7.900799&  
		7.997235& 
		7.90233 \\
		
		5.1.12& 
		7.902481& 
		7.902556 & 
		7.899145&  
		7.903374&  
		7.902974& 
		7.999652\\
		
		5.1.13& 
		7.902075& 
		7.902688 & 
		7.900901&  
		7.904566&  
		7.901951& 
		7.999612\\
		
		5.1.14& 
		7.902918& 
		7.903474 & 
		7.900112& 
		7.903111& 
		7.902577& 
		7.901996 \\
		
		5.2.08& 
		7.903094& 
		7.903953 & 
		7.902325& 
		7.901762& 
		7.903408& 
		7.999608 \\
		
		5.2.09& 
		7.902541& 
		7.902233 & 
		7.902001& 
		7.905854&  
		7.997842& 
		7.902996 \\
		
		5.2.10& 
		7.902029& 
		7.900714 & 
		7.902721& 
		7.902768& 
		7.997485& 
		7.999699 \\
		
		5.3.01& 
		7.902361& 
		7.902727 & 
		7.902432&  
		7.901040&  
		7.996482& 
		7.999658 \\
		
		5.3.02& 
		7.903230& 
		7.903182 & 
		7.902631&  
		7.900981& 
		7.903331& 
		7.90421 \\
		
		7.1.01& 
		7.901931& 
		7.902173 & 
		7.902002& 
		7.902145& 
		7.992482& 
		7.999615 \\
		
		7.1.02&  
		7.902419& 
		7.900879 & 
		7.902821&  
		7.902157&  
		7.997842& 
		7.999678 \\
		
		7.1.03& 
		7.902170& 
		7.902543 & 
		7.902325& 
		7.900645& 
		7.991748& 
		7.901996 \\
		
		7.1.04& 
		7.903219& 
		7.901126 & 
		7.902411& 
		7.904141& 
		7.992748& 
		7.902539 \\
		
		7.1.05& 
		7.902091& 
		7.903579 & 
		7.902251& 
		7.900027&  
		7.996748& 
		7.902605 \\
		
		7.1.06&  
		7.902850& 
		7.901930 & 
		7.902762&  
		7.901736& 
		7.902012& 
		7.902311 \\
		
		7.1.07& 
		7.902258& 
		7.903000 & 
		7.902575& 
		7.900802&  
		7.997485& 
		7.902568 \\
		
		7.1.08& 
		7.902022& 
		7.903197 & 
		7.902114&  
		7.900944& 
		7.902748& 
		7.902512 \\
		
		7.1.09& 
		7.902255& 
		7.902308 & 
		7.902709& 
		7.905658& 
		7.997485& 
		7.901951 \\
		
		7.1.10& 
		7.902032& 
		7.899542 & 
		7.902525&  
		7.893848& 
		7.991748& 
		7.903225 \\
		
		7.2.01&  
		7.902038& 
		7.902772 & 
		7.902224&  
		7.904525& 
		7.995748& 
		7.999652 \\
		
		Boat.512& 
		7.901863& 
		7.901908 & 
		7.902616& 
		7.900712&  
		7.992748& 
		7.999656 \\
		
		Gray21.512& 
		7.902807& 
		7.900170 & 
		7.902020& 
		7.902149& 
		7.993748& 
		7.999655 \\
		
		House & 
		7.90228701& 
		7.903580& 
		7.904501&  
		7.902156&  
		7.998748& 
		7.999665 \\
		
		Ruler.512& 
		7.901977& 
		7.903265 & 
		7.902454&  
		7.901428& 
		7.996748& 
		7.999678 \\
		
		Numbers.512& 
		7.903047& 
		7.903615 & 
		7.902535& 
		7.903579& 
		7.997748& 
		7.999675 \\
		
		Mean & 
		7.902391  & 
		7.902381  & 
		7.903141  & 
		7.902128  & 
		7.992748  & 
		7.999675  \\
		\hline\hline
		Pass Rate& 
		27/27& 
		17/27& 
		22/27&  
		10/27& 
		7/27& 
		11/27 \\
		\hline
	\end{tabular}
\end{center}
	\label{tab1}
	\begin{flushleft}
		h$_{left}=$7.901515698 \par h$_{right}=$7.903422936
	\end{flushleft}
\end{table*}

\begin{table*}
	\caption{Randomness test of s-box and Encrypted image}
		\begin{center}
			\setlength{\extrarowheight}{3pt}
			\begin{tabular}{|p{150pt}|c c|c|}
				\hline
				\raisebox{-1.50ex}[0cm][0cm]{Test name}& 
				\multicolumn{2}{|c|}{P-value(for randomness $\ge $0.01)} & 
				\raisebox{-1.50ex}[0cm][0cm]{Result } \\
				\cline{2-3} 
				& 
				S-box& 
				Encrypted Image& 
				\\
				\hline
				Block frequency test& 
				0.2442& 
				0.4071& 
				Pass \\
				\hline
				The Runs Test, & 
				0.7493& 
				0.6758& 
				Pass
				\\
				\hline
				The Longest-Run-of-Ones & 
				0.2465& 
				0.4465& 
				Pass \\
				\hline
				The Binary Matrix Rank Test, & 
				0.4312& 
				0.4312& 
				Pass \\
				\hline
				The Discrete Fourier Transform& 
				0.0621& 
				0.6555& 
				Pass \\
				\hline
				The Cumulative Sums & 
				0.5283& 
				0.6170& 
				Pass \\
				\hline
				The Approximate Entropy Test & 
				0.6825& 
				0.6892& 
				Pass \\
				\hline
				The Non-overlapping Template & 
				0.7784& 
				0.7204& 
				Pass \\
				\hline
				The Overlapping Template & 
				0.4432& 
				0.5592& 
				Pass \\
				\hline
				The Linear Complexity Test & 
				0.5517& 
				0.5175& 
				Pass \\
				\hline
				The Serial Test & 
				P-value1$=$0.9070; \par P-value2$=$0.8547& 
				P-value1$=$0.8879; \par P-value2$=$0.8041& 
				Pass \\
				\hline
			\end{tabular}
			\label{tab8}
		\end{center}
\end{table*}
\begin{table*}
	\caption{The computational complexity and the time complexity using different algorithms }
	\setlength{\extrarowheight}{3pt}
\begin{center}
		\begin{tabular}{ c c c } 
		\hline
		Algorithm  & Computational complexity  & Execution time \\
		\hline \hline
		LICM \cite{cao2018novel}& $m\log(8n)+ 8n\log(m)+2(m +8n)$& 1.2395278 \\
		
		ICMIE \cite{cao2020designing}  &  $2(mlog(n)+mn)$ & 1.1858623 \\ 
		2D-LCCCM\cite{Nan2022} & $(5+CR)\times
		m\times n$ & 1.1256258\\ 	
		Cross-coupled chaos \cite{Patro2020} & $2\max(m,n)+9(m+n)$ & 1.1217369\\
		Mixed image element and chaos \cite{Zhang2017}& $\frac{(m\log(n)+n\log(m))mn}{64}$ & 2.1235896\\
		Our Scheme & $\frac{64\times m\times n}{k}+8(10m+17n)$ &1.1019249\\
		\hline
	\end{tabular}
\end{center}
	\label{tabtime}
\end{table*} 
\subsection{Local Shannon Entropy}
The Local Shannon entropy, which quantitatively measures the distribution of information, is used to estimate  the randomness of an encrypted image.  Mathematically, it is defined as 
\begin{equation}
\overline{H_{k,T_B}}=\sum_{i=1}^{k}\frac{H(s_i)}{k}
\end{equation}
where $s_1,s_2,...,s_k$ are $ k $ selected blocks with $T_B$ pixels of a chosen image. If $\overline{H_{k,T_B}}$  is in the interval of $(h_{left},h_{right})$, then the cipher text image will be considered as passing the test. Next,  we calculate the critical values $h_{left}$ and $h_{right}$ with  $\alpha$ level of significance in a $ Z-test$ as follows: 
\begin{equation}
\begin{split}
h_{left}= \mu_{\overline{H_{k,T_B}}} - \Phi_{\alpha/2}^{-1}\sigma_{\overline{H_{k,T_B}}}\\
h_{right}= \mu_{\overline{H_{k,T_B}}} + \Phi_{\alpha/2}^{-1}\sigma_{\overline{H_{k,T_B}}}
\end{split}
\end{equation}
where $\Phi^{-1}$ is the inverse cumulative density function of the standard normal distribution $N(0,1)$ and $\mu_{\overline{H_{k,T_B}}}$, and $\sigma_{\overline{H_{k,T_B}}}$ be the mean and variance of Local Shannon entropy calculated in $ k $ non overlapping blocks.

In this test, we select $27$ different images from USC-SIPI Miscellaneous data set, and then encrypt these images by various image encryption algorithms. According to the recommendation
in \cite{Saveriades2013}, we set the parameters $ (k, TB) = (30, 1936)$ and $\alpha=0.001$, then an image is considered to pass the test if the obtained Local Shannon Entropy falls into the interval $ (7.901515698, 7.903422936) $. Table \ref{tab1} lists Local Shannon entropy results of the encrypted image. The results demonstrate that the proposed algorithm successfully pass the test. 

\subsection{Randomness test for S-box and encrypted image}
To test the randomness of the proposed S-box and the encrypted image, we use statistical tests package, namely, NIST-800-22. This package has several tests, and each test provides a p-value, which can discover the non-random regions from several sides. If the p-value $ \geq $ $ 0.01 $, then the truncated sequence passes the test. Table\ref{tab8} lists NIST-800-22 results for the S-box and encrypted image. They successfully pass all the tests. That means the S-box and the pixel value of the encrypted image are random.

\subsection{Computational  and time complexity analyses}
{In the row-column permutation or the  confusion stage, the computational complexities to perform the row shuffling and the column shuffling  operations using the chaotic sequence, respectively, are $\textbf{O} (m)$  and $\textbf{O} (n)$, where $m$ and $n$ are the pixel values. So, the computational complexity to perform the pixel shuffling is $8(m+8n)$.} 
\par {In the diffusion stage, the images are first divided  into scramble matrices $SC^i_{k\times k}$ for $k=1, 2,..., 16$  and $i=1, 2,..., \frac{m\times n}{k^2}$. This means that   each block is of the order of $k\times k$ in which each vector is multiplied by the field matrix  of order $8\times 8$.  Thus, the time complexity is given as  $\textbf{O} (\frac{64\times m\times n}{k^2})$. Also,  in the final encryption in which bitwise-XOR   is performed, the time complexity is given as $72(m+n)$.  So, the total time complexity is   $\frac{64\times m\times n}{k^2}+8(10m+17n)$. As an illustration,   we consider a Barbara image of size $512\times 512$,   encrypt it $60$ times using R2016a Matlab software with i3-4005 CPU \@ 1.7 GHz, 4-Gb RAM, finally compare  with different existing schemes to show the efficiency of our proposed algorithm.  The results are dilplayed in Table \ref{tabtime}. }

\section{Conclusion}
\label{section:section7}
A low-dimensional discrete chaotic system such as the Duffing map can exhibit complicated multistability behaviors. The  coexistence of chaotic attractors with periodic orbits as well as  the coexistence of two chaotic attractors in the 2D Duffing map are shown. We have introduced the Sine-Cosine chaotification technique to enhance chaos complexity in the multistable regions of the 2D Duffing map. The proposed chaotification technique can be easily generalized to other low-dimensional chaotic maps. Several performance evaluations including the trajectory, Lyapunov exponents, bifurcations, FIPS 140-2 test, and Sample entropy have demonstrated that the enhanced Duffing map exhibits a wide hyperchaotic range, high randomness and extreme unpredictability.   Furthermore, its hyperchaotic sequences appear in a large area in the 2D phase space without exhibiting periodic behaviors. Consequently, the enhanced Duffing map could be a better choice than other existing chaotic maps for cryptography applications. Thus, we propose an  image encryption algorithm, which achieves the confusion and diffusion processes by hyperchaotic sequences, elliptic curve, and S-box. Simulation  results have revealed that the proposed encryption algorithm can give  the users a flexibility to encrypt several kinds of images such as Grey scale, Medical, and RGB images with a higher level of security. As the proposed image encryption algorithm based on the enhanced Duffing map has high security and efficiency, our future work will investigate its application in video encryption. \\

\section*{Compliance with ethical standards}

\textbf{Conflict of interest} All authors declare that they have no conflict of
interest.\\

\noindent\textbf{Ethical approval} This article does not contain any studies with human
participants or animals performed by any of the authors.\\

\noindent\textbf{Informed Consent} Not applicable.\\

\noindent\textbf{Author Contributions} Hayder Natiq conceived and designed
the analysis, collected the data, performed the analysis, wrote the paper;
Animesh Roy conceived and designed the analysis, wrote the paper; conceptualization and supervision, Santo Banerjee; methodology, A. P. Misra; software, N. A. A. Fataf.

\end{document}